\documentclass[times,twocolumn,final,authoryear]{elsarticle}

\usepackage{jasr}
\usepackage{framed,multirow}

\usepackage{amssymb}
\usepackage{latexsym}
\usepackage{amsmath,bm}

\usepackage[switch]{lineno}

\usepackage{url}
\usepackage{xcolor}
\definecolor{newcolor}{rgb}{.8,.349,.1}

\usepackage[citebordercolor=white]{hyperref}

\journal{Advances in Space Research}

\begin{document}

\verso{Hirdesh Kumar \textit{etal}}

\begin{frontmatter}

\title{On the propagation of gravity waves in the lower solar
atmosphere in different magnetic configurations}%

\author[1,2]{Hirdesh Kumar\corref{cor1}}
\cortext[cor1]{Corresponding author: 
  Email: hirdesh@prl.res.in}
\author[1]{Brajesh Kumar}
\author[3,4]{S. P. Rajaguru}

\address[1]{Udaipur Solar Observatory, Physical Research Laboratory, Dewali, Badi Road, Udaipur 313004 Rajasthan, India}
\address[2]{Department of Physics, Indian Institute of Technology Gandhinagar, Gandhinagar 382 355 Gujarat, India}

\address[3]{Indian Institute of Astrophysics, Bangalore-34, India}

\address[4]{Solar Observatories Group, Department of Physics and W.W. Hansen Experimental Physics Lab, Stanford University, Stanford CA, 94305-4085 USA.}


\begin{abstract}

Gravity waves are generated by turbulent subsurface convection
overshooting or penetrating locally into a stably stratified medium.
While propagating energy upwards, their characteristic negative
phase shift over height is a well-recognized observational signature.
Since their first detailed observational detection and estimates of
energy content, a number of studies have explored their propagation
characteristics and interaction with magnetic fields and other waves
modes in the solar atmosphere. Here, we present a study of the
atmospheric gravity wave dispersion diagrams utilizing intensity
observations that cover photospheric to chromospheric heights over
different magnetic configurations of quiet-Sun (magnetic network regions),
a plage, and a sunspot as well as velocity observations within the
photospheric layer over a quiet and a sunspot region. In order to
investigate the propagation characteristics, we construct two-height
intensity - intensity and velocity- velocity cross-spectra and study
phase and coherence signals in the wavenumber - frequency dispersion
diagrams and their association with background magnetic fields. We find signatures of association between magnetic fields and much
reduced coherence and phase shifts over height from intensity-intensity
and velocity-velocity phase and coherence diagrams, both indicating
suppression/scattering of gravity waves by the magnetic fields. Our
results are consistent with the earlier numerical simulations, which indicate
that gravity waves are suppressed or scattered and reflected back into the
lower solar atmosphere in the presence of magnetic fields.

\end{abstract}


\end{frontmatter}


\section{Introduction}
\label{sec1}

Waves in a compressible stratified medium in the presence of a gravitational field, like the 
atmosphere of the Earth or the Sun, can be driven by both compressional and buoyancy forces, resulting in a rich spectrum of acoustic-gravity waves. In the lower solar atmosphere, such waves are well 
recognised as an agent of non-thermal energy transfer, especially through their interactions with 
and transformations by the highly structured magnetic fields that thread these layers. These waves 
are generated by turbulent convection within and near the top boundary layers of the convection zone 
\citep{1948ApJ...107....1S, 1952RSPSA.211..564L, 1967SoPh....2..385S, 1990ApJ...363..694G} and the 
acoustic part of the spectrum resonate to form $p$-modes in the interior of the Sun. In the solar 
atmosphere, the behaviour of waves become more complicated owing to the sharp fall in density and 
the preferred direction imposed by gravity in the fluid: the propagation characteristics are 
anisotropic, in general. In addition, the stratification of the atmosphere also imposes height-dependent cutoff 
frequencies below which gravity modified acoustic waves cannot propagate \citep{Mihalas1995} upward of the respective
heights. The internal or atmospheric gravity waves (IGWs) are generated by turbulent subsurface convection 
overshooting or penetrating locally into a stably stratified medium \citep{1967IAUS...28..429L}. 
This is a normal response generated by a gravitationally stratified medium to any perturbations from 
its equilibrium position, and buoyancy acting as a restoring force. \cite{1967IAUS...28..429L} suggested that the oscillations observed in the upper photosphere and 
lower chromosphere can be interpreted as gravity waves, and also 
that radiative damping of such gravity waves provides a mechanism of heating of the lower 
chromosphere. One of the interesting properties of IGWs is that, while transporting energy upward 
from the photosphere to higher layers, they show a characteristic downward phase propagation 
\citep{Lighthill1978}. These waves play an important role in the transportation of energy and 
momentum and mixing the material in the regions that they propagate in. For example, the IGWs in the solar radiative interior
and in other stars, despite no clear detection of their signatures at the solar surface, are theorized to play key roles in the 
mixing and transport of angular momentum. 
The investigation of \cite{1981ApJ...249..349M, 1982ApJ...263..386M} revealed that gravity waves can 
reach a maximum height of 900 - 1600 km in the solar atmosphere depending upon the energy flux 
carried by these waves, before nonlinearities lead to wave breaking. Following the above initial studies, the propagation characteristics of IGWs in the solar atmosphere have been 
examined utilizing velocity-velocity, intensity-intensity observations or simulations or both by 
a good number of authors 
\citep{1989A&A...213..423D,2001A&A...379.1052K,2003A&A...407..735R,2008ApJ...681L.125S, 
2011A&A...532A.111K,2014SoPh..289.3457N, 2017ApJ...835..148V, 2019ApJ...872..166V, 
2020A&A...633A.140V, 2021RSPTA.37900177V}. \cite{2001A&A...379.1052K}, utilizing the 1700 {\AA} and 1600 {\AA} intensity observations obtained from the Transition Region and Coronal Explorer (TRACE) instrument of a quiet region observed in the disk centre, identified gravity waves in the $k_{h}-\nu$ phase diagram (c.f. Figure 24, \cite{2001A&A...379.1052K}). Subsequently, \cite{2003A&A...407..735R} using the simultaneous UV intensities (1700 {\AA} and 1600 {\AA} intensity images) obtained from the TRACE also detected the gravity waves in $k_{h}-\nu$ phase diagram (c.f., Figure 3,  \cite{2003A&A...407..735R}). They have also found the signature of gravity waves in the $k_{h}-\nu$ phase diagram constructed from white light and 1700 {\AA} intensity images (c.f., Figure 5, \cite{2003A&A...407..735R}). Using $V - V$ observations along with 3D numerical simulations, 
\cite{2008ApJ...681L.125S} detected the upward propagating atmospheric gravity waves. They also 
estimated that the energy flux carried by gravity waves was comparable to the radiative losses of the 
entire chromosphere. Using observations of Fe I 5576 {\AA} 
and Fe I 5434 {\AA} lines, \cite{2011A&A...532A.111K} studied acoustic and atmospheric gravity waves in the quiet Sun and 
estimated their energy transport to the chromosphere. They concluded that gravity waves also 
contribute in the chromospheric heating. Using multi-height velocity extractions from the Fe I 6173 {\AA} line filtergrams
provided by the HMI/SDO, \cite{2014SoPh..289.3457N} also reported the presence of atmospheric gravity waves in the 
solar atmosphere. More recently, using realistic 3D numerical simulations of the solar atmosphere to investigate the 
propagation dynamics of acoustic-gravity waves, \cite{2017ApJ...835..148V} conclude that IGWs are 
absent or partially reflected back into the lower layers in the presence of the magnetic fields. 
They further argue that the suppression is due to the coupling of IGWs to slow magnetoacoustic waves 
still within the high plasma-$\beta$ region of the upper photosphere. \cite{2019ApJ...872..166V} found that 
the propagation properties of IGWs depend on the average magnetic field strength in the upper 
photosphere and therefore these waves can be potential candidates for magnetic field diagnostics of 
these layers.

 In this article, we report our results from a detailed study of the atmospheric gravity wave 
dispersion diagrams utilizing intensity observations that cover photospheric to chromospheric 
heights over regions of different magnetic configurations: quiet-Sun (magnetic network regions), 
plage, and sunspot. Additionally, we have used two-height velocities estimated within Fe I 6173 {\AA} line over a quiet and a sunspot region. In order to investigate the propagation characteristics, we construct 
two-height intensity - intensity and velocity - velocity cross-spectra and study the phase and coherence 
signals in the wavenumber - frequency ($k_{h} - \nu$) dispersion diagrams and their association with background magnetic fields, 
utilizing long duration data sets situated at different locations on the solar disc.
 We compare the derived signatures of the interaction between the IGWs and magnetic fields with those 
reported using numerical simulations by \cite{2017ApJ...835..148V, 2019ApJ...872..166V}, and \cite{2020A&A...633A.140V}. The article is structured as follows: Section 2 discusses the observations, 
data sets, and analysis procedures, Section 3 presents the results obtained in this investigation, 
followed by Section 4 that presents discussions and conclusions.

\section{Observational Data and Analyses}

\begin{figure*}[ht!]
\centering
\includegraphics[scale=0.365]{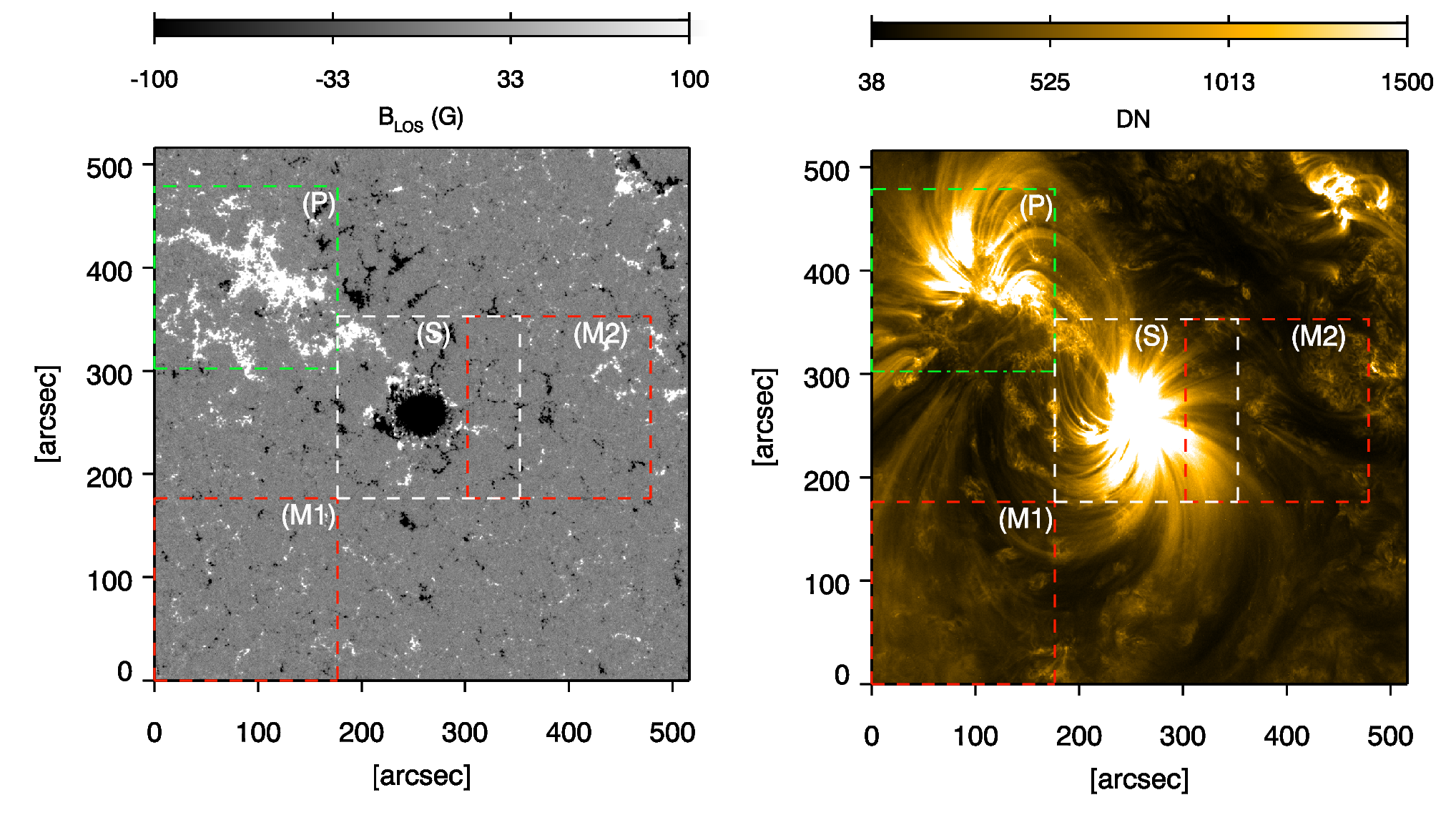}
\caption{\textit{Left panel}: HMI line-of-sight magnetic field of a large region observed on August 3, 2010 corresponding to the start time of data used in this work. The colored dashed regions mark the boundaries of sub-regions studied in this work: regions enclosed in red, white and green colored boxes mark the quiet magnetic network, sunspot and plage regions, and also denoted by $M1$, $M2$, $S$, $P$,  respectively. The magnetic field grey scale has been saturated at $\pm$ 100 G to view better the small scale magnetic fields. \textit{Right panel}: Same as left panel but from AIA 171 {\AA} channel. 
\label{fig:hmiblos}}
\end{figure*}

\begin{figure*}[ht!]
\centering
\includegraphics[scale=0.365]{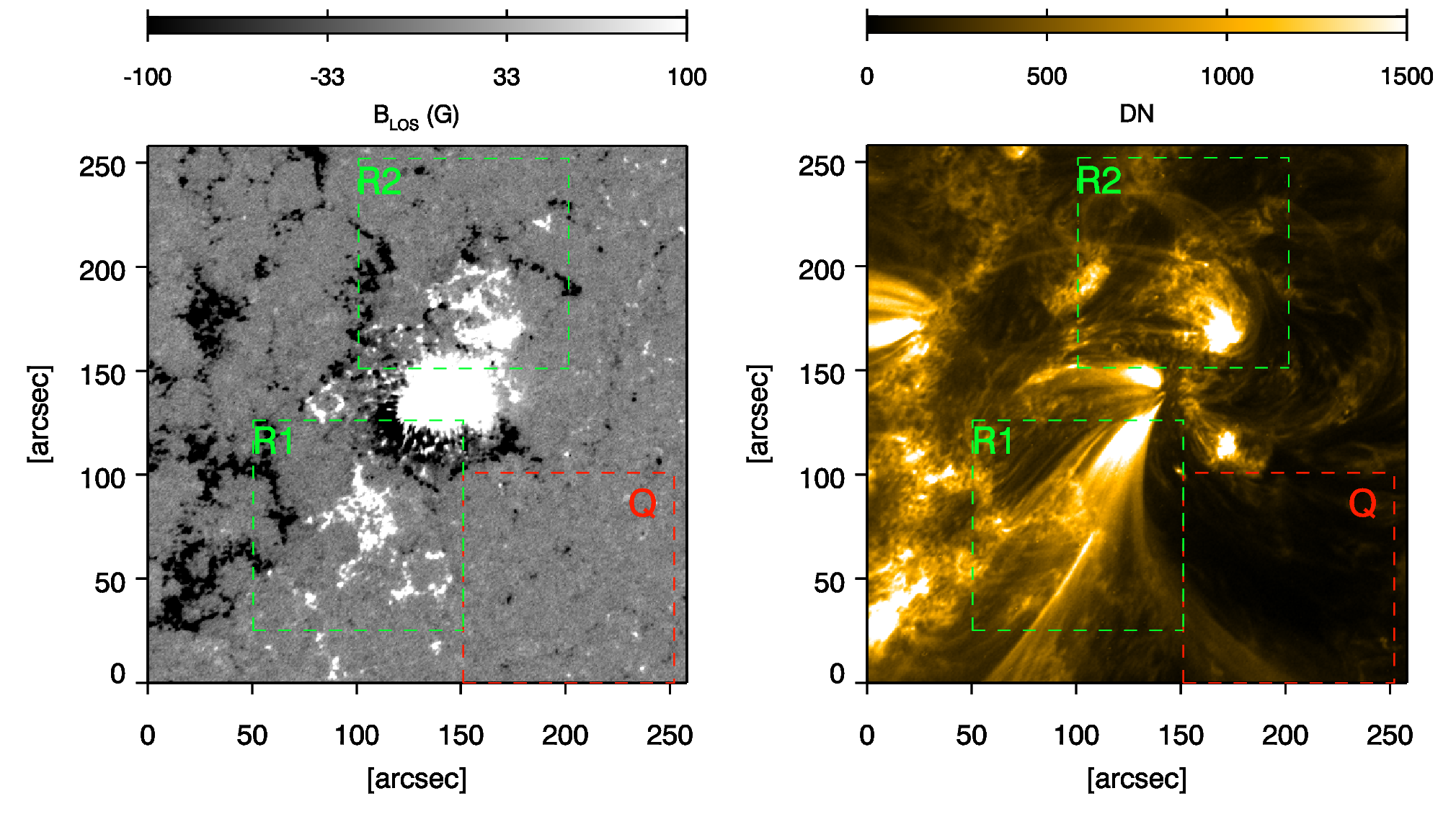}
\caption{\textit{Left panel}: HMI line-of-sight magnetic field of a large region observed on October 11, 2014 corresponding to the start time of data used in this work. The colored dashed regions mark the boundaries of sub-regions studied in this work: regions enclosed in red, and green colored boxes mark the quiet magnetic network, canopy regions, and also denoted by $Q$, $R1$, and $R2$,  respectively. The magnetic field grey scale has been saturated at $\pm$ 100 G to view better the small scale magnetic fields. \textit{Right panel}: Same as left panel but from AIA 171 {\AA} channel. 
\label{fig:hmiblos_dataset2}}
\end{figure*}

We employ two-height cross-spectra of intensities and velocities observed over regions of interest to study the height evolution of wave phases and interaction with the background magnetic fields. Intensity - 
intensity ($I - I$) cross spectra utilise observations obtained from the Helioseismic and Magnetic 
Imager (HMI; \citet{2012SoPh..275..229S} instrument onboard the Solar Dynamics Observatory (SDO; 
\citet{2012SoPh..275....3P}) for the photosphere using the Fe I 6173 {\AA} line, and from the
Atmospheric Imaging Assembly (AIA; \cite{2012SoPh..275...17L}) onboard the SDO for the 1700 {\AA} and 
1600 {\AA} UV channels that sample the photosphere - chromosphere.   Velocity - Velocity ($V - V$) cross spectra utilise observation obtained using photospheric Fe I 6173 {\AA} line from the Interferometric BI-dimensional Spectrometer (IBIS; \cite{2006SoPh..236..415C}) instrument installed at the Dunn Solar Telescope (DST) at Sacramento Peak, New Mexico. We analyze observations of three large regions, identified as data set 1, 2, and 3, of widely differing magnetic configurations. These regions are also situated on the different locations on the Sun. The data set 1 covers quiet or weak magnetic patches identified by red dashed boxes and labelled $M1$ and $M2$, a plage and sunspot (NOAA AR 11092)  areas identified by green and 
white dashed boxes labelled $P$, $S$, respectively (c.f., Figure \ref{fig:hmiblos}). This whole region was situated near 
the disk centre (63$''$,150$''$ solar coordinates) and was observed 
on August 03, 2010. The observables include photospheric line-of-sight magnetograms ($B_{LOS}$), continuum 
intensity ($I_{c}$), UV intensities in 1700 ($I_{uv1}$) and 1600 ($I_{uv2}$) {\AA} filters, and the data 
have been tracked and remapped for a duration of 14 hours; this data set has previously been used for a study of 
high-frequency acoustic halos by \cite{2013SoPh..287..107R} and also for a study of the 
propagation of low-frequency acoustic waves into the chromosphere by \cite{2019ApJ...871..155R}. The 
data set 2 includes a quiet patch identified by red dashed square box labelled $Q$, and two further sub-regions near 
the sunspot (NOAA AR 12186) demarcated by green square boxes labelled $R1$ and $R2$ (c.f. Figure \ref{fig:hmiblos_dataset2}).
This large region was away from the disk centre (-309$''$,-418$''$ solar coordinates) and was observed on October 11, 2014, and the same observables as for data set 1 are tracked and remapped for a duration of 12 h 47 minutes.  The data set 3 covers a quiet and a sunspot (NOAA AR 10960) region demarcated as black square box and labelled $A$ and $B$ (c.f. Figure \ref{IBIS_Sample_Image}), which are identified from the  field-of-view of the IBIS instrument. The observed region was also located close to the disk centre (276$''$,-227$''$ solar coordinates) on June 8, 2007. The identified regions (quiet ($A$) and sunspot ($B$)) comprise 3 hours 44 minutes duration and utilized two height velocities within Fe I 6173 {\AA} line formation region. This IBIS data set has been previously used for various studies \citep{2010ApJ...721L..86R, 2012SoPh..278..217C, 2022ApJ...933..109Z}. The values of average of absolute line-of-sight magnetic fields integrated over whole observation period in the $M1$, $M2$, $P$, $S$, $Q$, $R1$, $R2$, $A$, and $B$ region are tabulated in Table \ref{AVG_BLOS_TABLE}. Brief descriptions of the instruments and data 
reductions done for the above observations are provided in the following sub-sections.

\begin{table*}
\centering
\caption{Table represents the values of average of absolute line-of-sight magnetic fields over selected locations.} 
\begin{tabular}{|p{3cm}|p{3cm}|}
 \hline
 Location& $<|B_{LOS}|>$ (G)\\
 \hline
 $M1$   & 3.3   \\
 $M2$ &   5.5  \\
 $P$ & 37.0 \\
 $S$    & 53.2 \\
 $Q$ &   2.7 \\
 $R1$ &   32.2 \\
 $R2$ &   32.8 \\
 $A$  &   5.8\\
 $B$  &   74.2\\ 
 
  \hline
\end{tabular}
\label{AVG_BLOS_TABLE}
\end{table*}

\subsection{HMI and AIA Observations}

The HMI instrument observes the photosphere of the Sun in Fe I 6173 {\AA} absorption spectral line 
at six different wavelength positions. It provides full disk (4096$\times$4096 pixel$^{2}$) 
continuum intensity, line depth, Dopplergrams, and line-of-sight magnetograms at a spatial sampling 
of 0.504$''$ per pixel and at a temporal cadence of 45 s. It also provides full-disk vector 
magnetic field at a slightly lower cadence (12 minutes) as a standard product, although it can also 
be obtained at a cadence of 135 s \citep{2014SoPh..289.3483H}. The AIA instrument observes the outer 
atmosphere of the Sun in seven extreme ultraviolet (EUV) filters at 94, 131, 171, 193, 211, 304 and 
335 {\AA}, two ultraviolet (UV) filters at 1600 and 1700 {\AA} and one white light filter at 4500 
{\AA} wavelengths. It provides full disc images (4096$\times$4096 pixel$^{2}$) at a spatial sampling 
of 0.6$''$ per pixel. The temporal cadence is 12 s for EUV, 24 s for UV, and 3600 s for the 
white light filter. Here, we have used photospheric continuum intensity ($I_{c}$), and line-of-sight 
magnetograms ($B_{LOS}$) from the HMI at a cadence of 45 s and UV observations in 1700 ($I_{uv1}$) and 
1600 ($I_{uv2}$) {\AA} at a cadence of 24 s obtained from the AIA instrument, respectively. We have 
tracked and remapped the images of all these observables to the same spatial and temporal sampling 
as HMI observations. The sub-regions $M1$, $M2$, $P$, and $S$ from data set 1 are of size 176$\times$176 arcsec$^2$, as shown in the Figure \ref{fig:hmiblos}, gives us a wavenumber resolution ($\Delta k_{h}$) of 0.049 rad Mm$^{-1}$. The total time duration of data set 1 is 14 hours and cadence (45 s), gives us a frequency resolution ($\Delta \nu$) of 19.8 $\mu$Hz, and Nyquist frequency ($\nu_{Ny}$) of 11.11 mHz. The spatial resolution $\delta x$ = 0.504$''$ per pixel corresponds to Nyquist wavenumber ($k_{Ny} = \pi/\delta x$) of 8.55 rad Mm$^{-1}$. The data set 2 comprise of $Q$, $R1$, and $R2$ regions of 100$\times$100 arcsec$^{2}$ shown in the Figure \ref{fig:hmiblos_dataset2}, gives us a wavenumber resolution ($\Delta k_{h}$) of 0.0859 rad Mm$^{-1}$. The total time duration of data set 2 is 12 h 47 minutes, spatial sampling and temporal cadence are same as data set 1. Hence, it gives us a frequency resolution ($\Delta \nu$) of 21.7 $\mu$Hz.

\subsection{IBIS Observations} 

\begin{figure*}[ht!]
\centering
\includegraphics[width=0.47\textwidth]{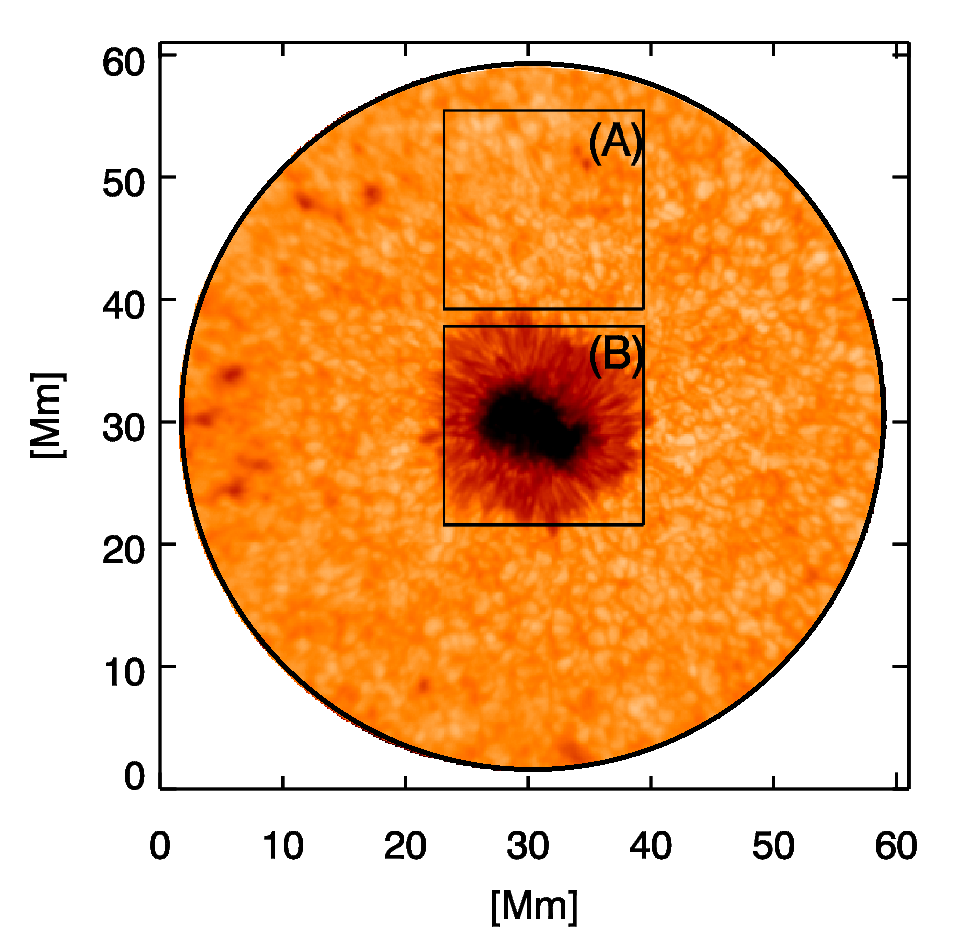}\hspace*{0.34cm}
\includegraphics[width=0.47\textwidth]{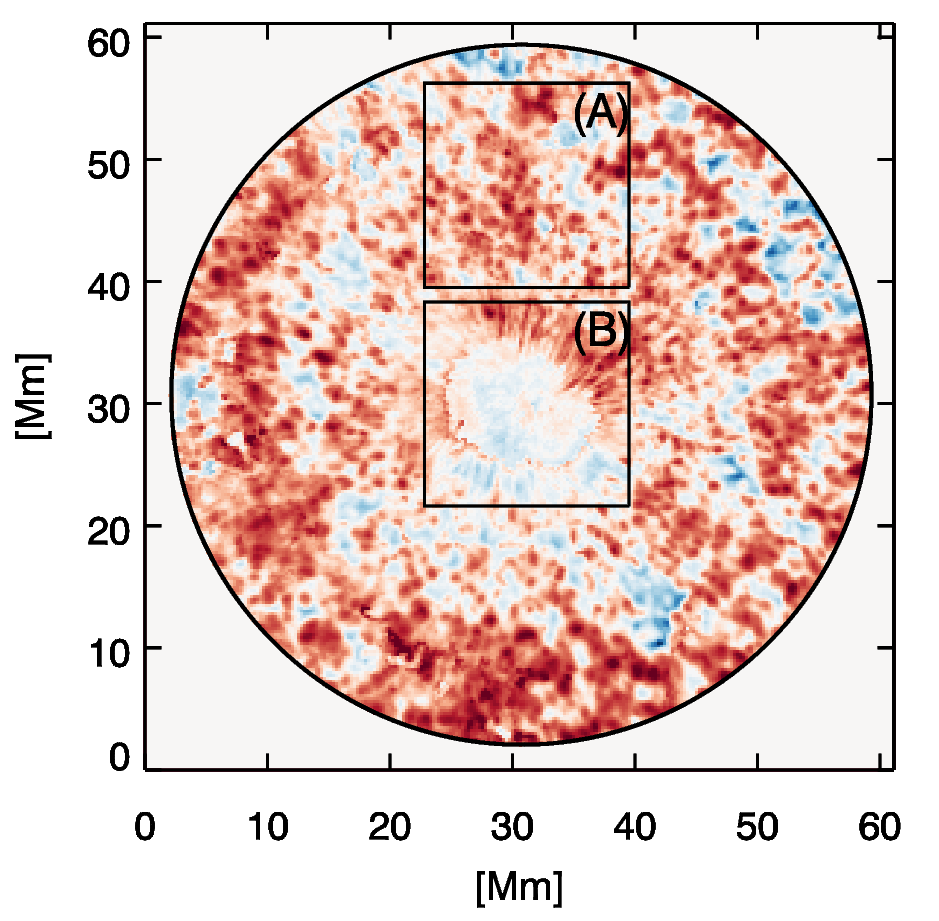}
\caption{{\textit{Left panel}: A sample image of continuum intensity showing the field of view of the observation obtained from the IBIS/DST with a sunspot at the center. The black square regions mark the boundaries of sub-regions of a quiet and a sunspot region studied in this work, also denoted by $A$, and $B$, respectively. \textit{Right panel}: A sample image of line-of-sight Doppler velocity estimated by bisector method at 10$\%$ intensity levels i.e. V$_{10}$ velocity map.}}
\label{IBIS_Sample_Image}
\end{figure*}

 The imaging spectropolarimetry data were acquired on June 08, 2007 utilizing the IBIS instrument installed at the DST. The IBIS has a circular FOV of 80$''$ in diameter ($\approx$ 60 Mm). The observed region covers a medium size sunspot and surrounding quiet-Sun situated near the disk center, with full Stokes ($I$,$Q$,$U$,$V$) images scanned over 23 wavelength positions along the Fe I 6173.3 {\AA} line at a spectral resolution of 25 m{\AA}. The spatial resolution of these observations is 0.33$''$ (0.165$''$ per pixel) at a temporal cadence of 47.5 s. From these observations \cite{2010ApJ...721L..86R} have estimated 10 bisector line-of-sight Doppler velocities starting from line core (level 0) to line wing (level 9). Here, we have used two height velocities at 10$\%$ ($V_{10}$) and 80$\%$ ($V_{80}$) intensity levels in our analysis, in which $V_{80}$ corresponds to lower height, whereas $V_{10}$ corresponds to upper height within the Fe I 6173 {\AA} line. The data set 3 comprises of a quiet ($A$) and a sunspot ($B$) region of 24$\times$24 arcsec$^{2}$ as shown in Figure \ref{IBIS_Sample_Image}. The total time duration of data set 3 is 3 hours 44 minutes, and temporal cadence (47.5 s), gives us a frequency resolution ($\Delta \nu$) of 74.4 $\mu$Hz, and Nyquist frequency ($\nu_{Ny}$) of 10.5 mHz. The spatial resolution ${\delta x}$ = 0.165$''$ per pixel corresponds to Nyquist wavenumber ($k_{Ny} = \pi/\delta x$) of 26.2 rad Mm$^{-1}$.

\subsection{Height Coverage of HMI and AIA Observations}

The HMI and AIA observables, as described above, are chosen to cover the photospheric to mid-chromospheric layers
of the solar atmosphere. The formation height of Fe I 6173 {\AA} line in the solar photosphere was investigated by \cite{2006SoPh..239...69N},
who derived a height range of $h$ = 16 -- 302 km above the continuum optical depth at 5000 \AA ($\tau_{c}$ = 1, which corresponds to z = 0) in the quiet region. However, the formation height of Fe I 6173 {\AA} line in the magnetized region i.e. umbra is lower compare to quiet region. \cite{2006SoPh..239...69N} utilizing the Maltby-M umbral atmosphere model reported that core of Fe I 6173 {\AA} line form around 270 km, while wings form around 20 km above $\tau_{c}$ = 1 continuum optical depth. Non-LTE radiation hydrodynamic simulations done by \cite{2005ApJ...625..556F} show 
that the average formation heights of 360 and 430 km for the 1700 and 1600 {\AA} passbands, respectively, although
the 1600 {\AA} passband has contributions from a wider height-range extending into upper chromosphere.

\subsection{Cross-Spectral Analysis of Wave Propagation}

We investigate the cross-spectra of IGWs utilizing different observables covering heights from 
photosphere (20 km) to chromosphere (430 km). We form two-height intensity - intensity ($I - I$) and velocity - velocity ($V - V$) 
pairs and study their cross-spectra. Cross-spectra of two observables $f_{\rm{1}}(x,y,t)$ and 
$f_{\rm{2}}(x,y,t)$ are defined as the complex-valued product of their three dimensional Fourier 
transforms \citep{2017ApJ...835..148V},

\begin{equation}
    \label{eq:1}
	X_{12}(\bm k,\nu)= {\bm f}_{\rm{1}}(\bm k,\nu){\bm f^{*}}_{\rm{2}}(\bm k,\nu)
\end{equation}

where ${\bm f}$'s are the Fourier transforms, with a superscript $*$ representing the complex 
conjugate, ${\bm k}={\bm k_{h}}=(k_{x},k_{y})$ is the horizontal wave vector and $\nu$ is the 
cyclic frequency. Here, subscripts $1$ and $2$ in $f_{\rm{1}}$ and $f_{\rm{2}}$ denote the two 
heights $z_{1}$ and $z_{2}$ of the observables. The phase spectrum, $\delta\phi (\bm k,\nu)$, 
that captures the phase evolution between heights of the two observables is then given by the phase 
of the complex cross-spectrum $X_{12}(\bm k,\nu)$,

\begin{equation}
    \label{eq:2}
\delta\phi(\bm k,\nu) = \tan^{-1}[Im(X_{12}(\bm k,\nu))/Re(X_{12}(\bm k,\nu))]
\end{equation}

The normalised magnitude of $X_{12}(\bm k,\nu)$ is used to calculate the coherence,
\begin{equation}
    \label{eq:3}
	C(\bm k,\nu) = \frac{|X_{12}(\bm k,\nu)|}{\sqrt{|{\bm f}_{\rm{1}}|^{2}|{\bm f}_{\rm{2}}|^{2}}},
\end{equation}

 We focus on studying $\delta\phi (k_{h},\nu)$ from the different $I - I$ and $V - V$ cross-spectra
that trace photopsheric - chromospheric height ranges, viz. the HMI continuum $I_{c}$,
AIA UV intensities $I_{uv1}$ (1700 {\AA}) and $I_{uv2}$ (1600 {\AA}) offer pairs of heights from
among 20, 360, and 430 km, respectively and two height velocities estimated within Fe I 6173 {\AA} line formation region i.e. V$_{80}$ and V$_{10}$ are correspond to within 16 -- 300 km height range above {\bf $\tau_{c} = 1$}. In all our analyses, we azimuthally average the 
three-dimensional spectra in the $k_{x}-k_{y}$ plane to derive $k_{h}-\nu$ diagrams of 
$\delta\phi (k_{h},\nu)$ and $C(k_{h},\nu)$, where $k_{h}$ = $\sqrt{k_{x}^2+k_{y}^2}$.

To guide the analysis, we employ the well known dispersion relation, derived under the Cowling 
approximation that neglects perturbations in the gravitational potential for adiabatic 
acoustic-gravity waves in the solar interior and atmosphere \citep{1981NASSP.450..263L},

\begin{equation}
\centering
k_{z}^2 = \frac{(\omega^2-\omega^2_{ac})}{c^2_{s}} - \frac{(\omega^2-N^2)}{\omega^2}k^2_{h}
\label{dispr}
\end{equation}

where $\omega = 2\pi\nu$ is the angular frequency, $\omega_{ac}$ is the acoustic cutoff frequency and $N$ is 
the Brunt-V{\"a}is{\"a}l{\"a} frequency. In a $k_{h}-\nu$ diagram, regions where $k_{z}^2>0$ demarcates vertically propagating acoustic-gravity 
waves from the evanescent ($k_{z}^2<0$) ones. In all our two-height $\delta\phi (k_{h},\nu)$ that 
we derive from observations, we mark such propagation boundaries by evaluating the $k_{z}^2$ = 0 
condition using the dispersion relation given by Eqn. \ref{dispr} for the upper height using the 
VAL-C model of the solar chromosphere \citep{1981ApJS...45..635V}. The expressions for $\omega_{ac}$ 
and $N$ applicable for a gravitationally stratified atmospheres are,

\begin{equation}
\centering
\omega_{ac}^2 =\frac{c_{s}^2}{4H_{\rho}^2} (1 - 2\frac{dH_{\rho}}{dz})
\label{wac}
\end{equation}

\begin{equation}
\centering
N^2 = \frac{g}{H_{\rho}} - \frac{g^2}{c_{s}^2} 
        \label{bvfreq}
\end{equation}

We have two solutions for $k_{z}^2$ = 0, and hence two propagation boundaries separating vertically 
propagating waves from the evanescent ones in the $k_{h}-\nu$ diagram: the higher frequency 
boundary corresponds to the cut-off frequency for acoustic waves while the lower frequency one, 
typically falling lower than the $f$-mode frequencies, is that for the internal or atmospheric 
gravity waves. The locations of these boundaries in the $k_{h}-\nu$ plane depends on height in 
the atmosphere, and we typically overplot these boundaries for upper heights (solid black line) 
involved in each pair of variables for which the cross-spectral phases, ($\delta\phi$), coherences, 
($C$) are derived. We also overplot the dispersion curve of the surface gravity mode ($f$-mode) and 
that of acoustic Lamb mode in each diagrams.

\begin{figure*}[ht!]
\centering
\includegraphics[scale=0.26]{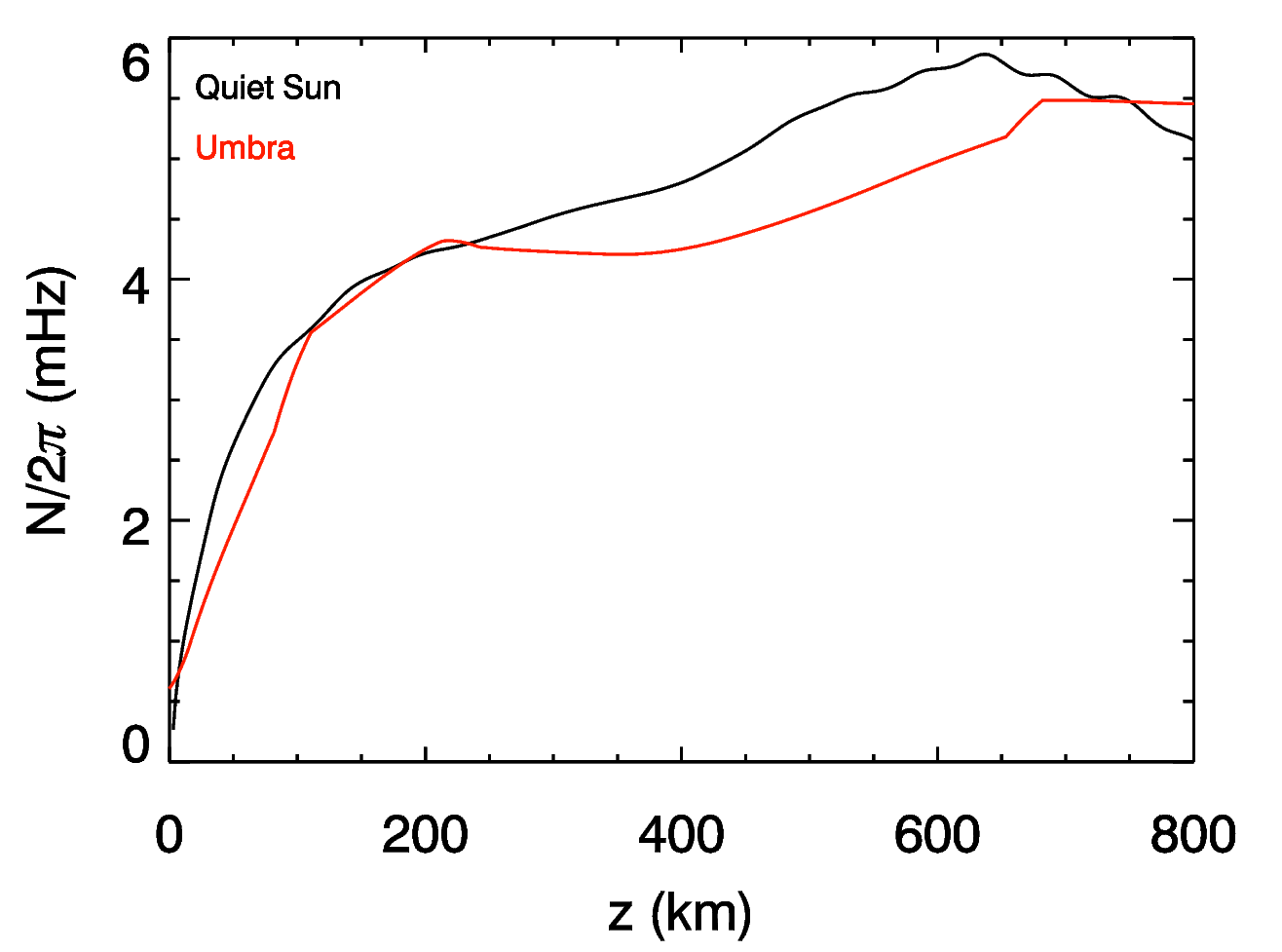}
\caption{Height evolution of Brunt-V{\"a}is{\"a}l{\"a} frequency ($N$) for quiet-Sun (black) of a realistic solar atmosphere using the VAL-C model and umbra (red) of a sunspot using Maltby-M model.}
\label{Plot_Brunt_Vaisal_Freq}
\end{figure*}

\begin{figure*}[ht!]
\centering
\includegraphics[scale=0.43]{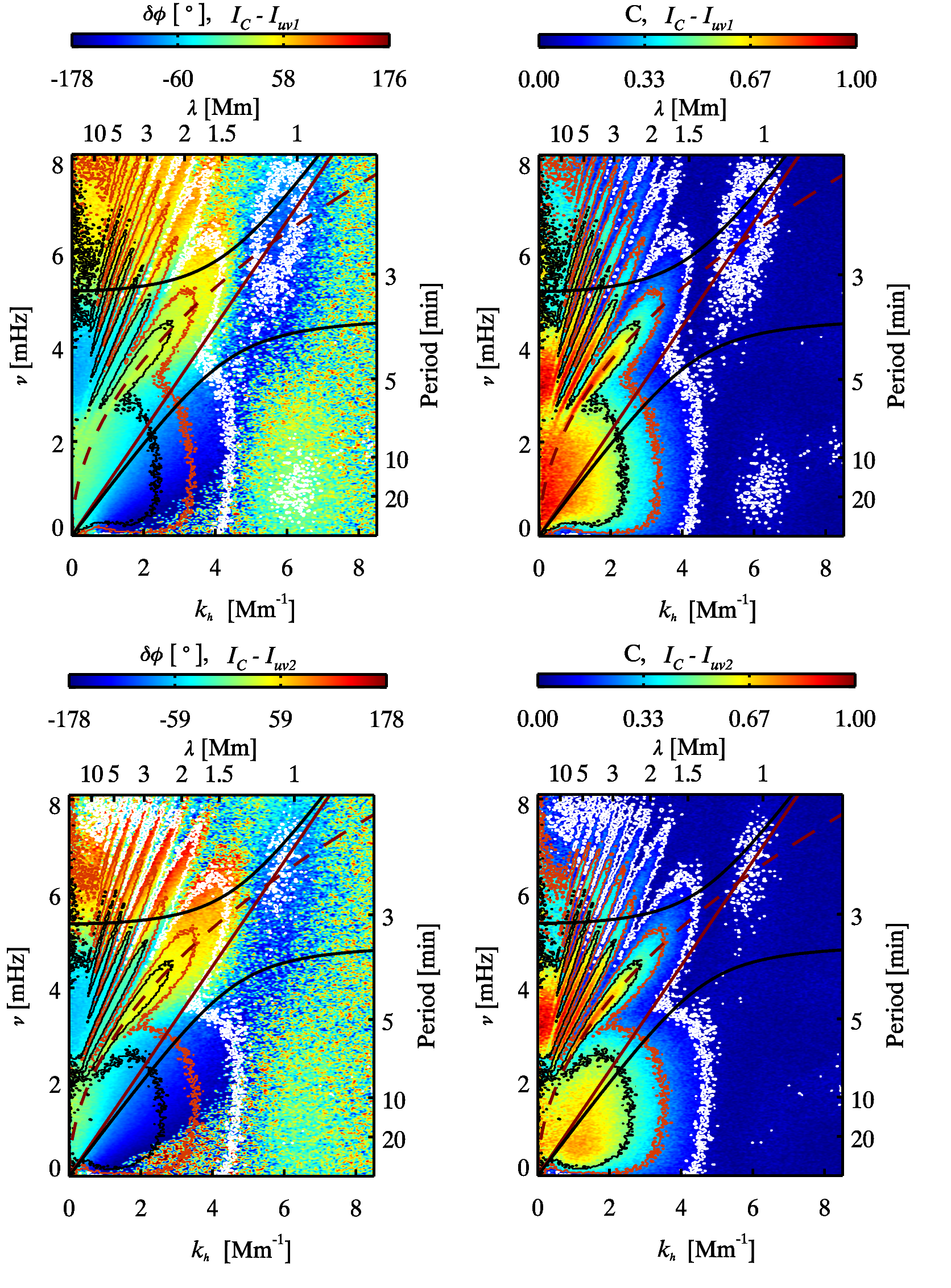}
\caption{\textit{Top panel}: Cross-spectral phase difference, $\delta\phi(k_{h},\nu)$ (\textit{top left panel}), and coherence, C$(k_{h},\nu)$ (\textit{top right panel}), diagrams of $M1$ region constructed from  $I_{c} - I_{uv1}$ pair of photospheric continuum intensity (HMI) and UV 1700 {\AA} channel of AIA, which correspond to 20 -- 360 km above z = 0 in the solar atmosphere. \textit{Bottom panel}: same as \textit{top panel}, but from $I_{c} - I_{uv2}$ pair of photospheric continuum intensity (HMI) and UV 1600 {\AA} channel of AIA, which  correspond to 20 -- 430 km above z = 0 in the solar atmosphere. The solid black lines separate vertically propagating waves ($k_{z}^2>0$) from the evanescent ones ($k_{z}^2<0$) at upper height. The dashed red line is the $f$-mode dispersion curve and solid red line is the Lamb mode. The overplotted black, red and white contours represent the coherence at 0.5, 0.3 and 0.1 levels, respectively.}
\label{M1PhaseCoherence}
\end{figure*}

\begin{figure*}[ht!]
\centering
\includegraphics[scale=0.43]{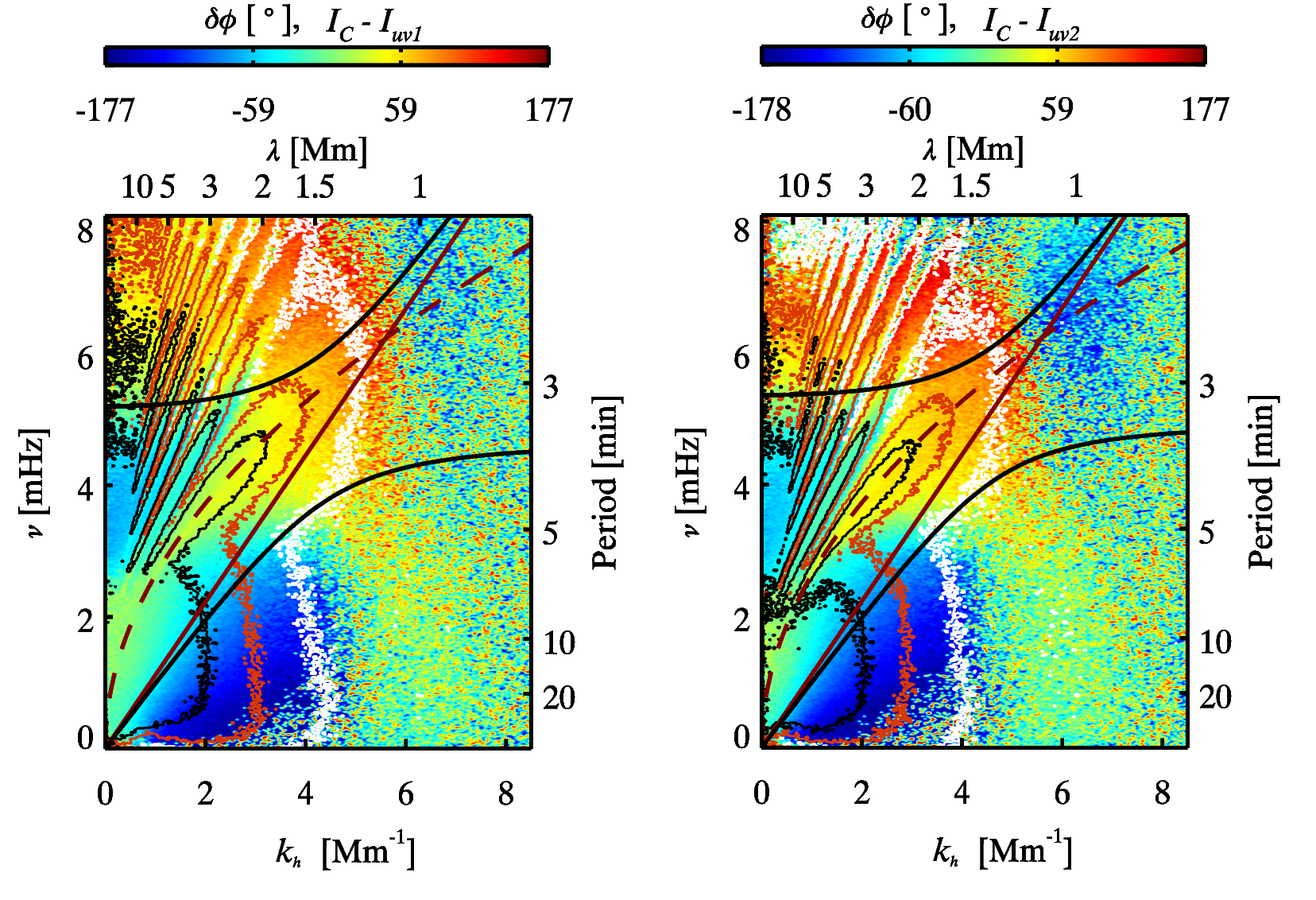}
\caption{{Cross-spectral phase difference, $\delta\phi(k_{h},\nu)$ diagrams of $M2$ region as indicated by $M2$ in Figure \ref{fig:hmiblos} constructed from  $I_{c} - I_{uv1}$, and  $I_{c} - I_{uv2}$ intensity pairs, respectively. The overplotted black, red and white contours represent the coherence at 0.5, 0.3 and 0.1 levels, respectively.}}
\label{M2PhaseCoherence}
\end{figure*}

\begin{figure*}[ht!]
\centering
\includegraphics[scale=0.43]{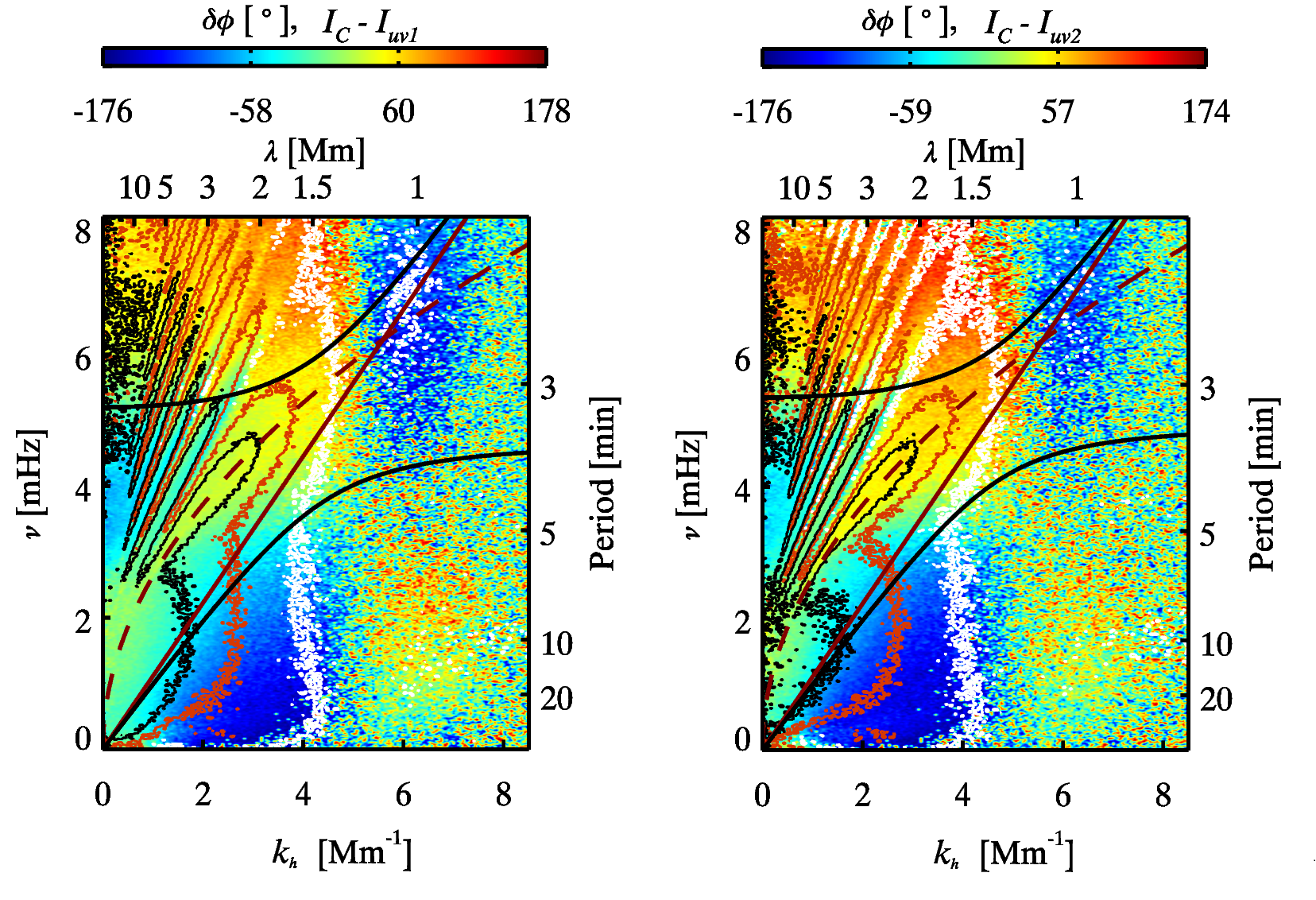}
\caption{{Same as Figure \ref{M2PhaseCoherence}, but for plage region as indicated by $P$ in Figure
\ref{fig:hmiblos}.}}
\label{Plage_PhaseCoherence}
\end{figure*}

\section{Results}

The IGWs in the solar atmosphere have their sources in the photospheric granular convection, which
overshoots into the stable layers above. As alluded to earlier, the IGWs have the characteristics of transporting energy 
upward while their phases propagate downward, hence exhibit negative phase in the gravity wave regime of $k_{h} - 
\nu$ cross-spectral phase diagram.  We first present and analyse the phase and coherence signals in the
$I - I$ cross-spectra that we obtain for the two large regions covering different magnetic
configurations, viz. quiet magnetic network, plage, and sunspots, which are shown in Figures \ref{fig:hmiblos} and
\ref{fig:hmiblos_dataset2}. Next, we analyse the phase and coherence diagrams constructed from $V - V$ pair over a quiet and a sunspot region. The wave propagation boundary corresponding to $k_{z}^2$ = 0 for the
IGWs is set by the Brunt-V{\"a}is{\"a}l{\"a} frequency, $N$, which is a function of height in the solar 
atmosphere and is given by equation \ref{bvfreq}. Using the VAL-C model \citep{1981ApJS...45..635V} for quiet Sun and the Maltby-M model \citep{1986ApJ...306..284M} for sunspot umbra, we have calculated $N$ and plotted in Figure \ref{Plot_Brunt_Vaisal_Freq}. It shows that $N$ increases upto a height of $\approx$ 600 km and then it starts decreasing, and it also indicates that there is no significant difference between the variation of $N$ in the quiet-Sun and umbra of a sunspot. For the mean height of formation of intensity $I_{uv2}$, 
$N$ approaches $\approx$ 5 mHz. \\

As the primary objective of this work is to study the effect of
magnetic fields on the propagation of IGWs, we first discuss the phase spectra of gravity waves in the quiet magnetic network regions followed by a comparision between the phase obtained for strongly magnetized regions and quiet regions. Results on effects of magnetic fields on the coherence of IGWs are presented at the end.

\begin{figure*}[ht!]
\centering
\includegraphics[scale=0.42]{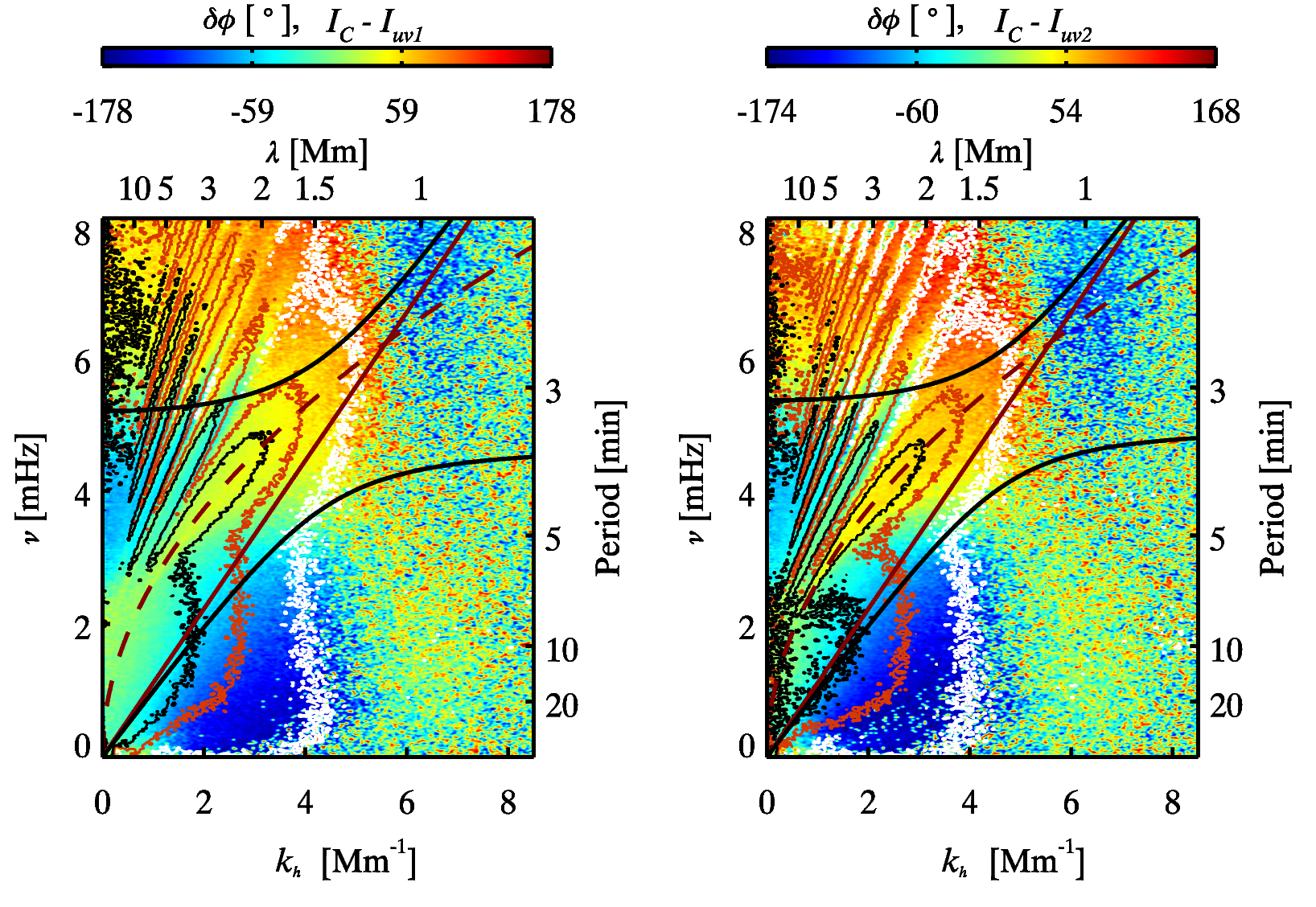}
\caption{Same as Figure \ref{M2PhaseCoherence} but for sunspot region as indicated by symbol $S$ in
the Figure \ref{fig:hmiblos}.}
\label{Sunspot_PhaseCoherence}
\end{figure*}

\subsection{Phase Spectra of IGWs in Quiet-Sun Regions}

The region labelled $M1$ in Figure \ref{fig:hmiblos} has the average absolute LOS magnetic field of $\approx$ 3 G (integrated over whole observation period) and is
chosen to represent the quiet-Sun. The cross-spectral phase difference, $\delta\phi$, and coherence, $C$, diagrams 
obtained for this region (from data set 1) are shown in Figure \ref{M1PhaseCoherence}: top panels are for the 
$I_{c}-I_{uv1}$ pair, and bottom panels are for the $I_{c}-I_{uv2}$ pair.

We mainly focus on the effect of magnetic fields on the gravity waves, and hence
are not discussing the acoustic wave regimes. In the gravity wave region of the 
diagnostic diagrams (c.f. Figure \ref{M1PhaseCoherence}), we observe the well known negative phase
as expected for IGWs, whose phase and group velocities have 
opposite signs. In general, in all our $\delta\phi(k_{h},\nu)$ diagrams, we observe noisy phase signals
with very low coherence $C$ for $k_{h} >$ 5.5 Mm$^{-1}$ and also even at lower $k_{h}$ when $\nu <$ 0.5 mHz.
We avoid such regions in $(k_{h},\nu)$ and focus only on IGWs that have $\nu \geq$ 0.5 mHz and $k_{h} <$ 5.5 Mm$^{-1}$. The magnitudes and wavenumber-extents of IGWs depend on the height separation in the solar atmosphere, and also
vary significantly from region to region or location even in the quiet-Sun.
For the quiet-Sun region $M1$, interestingly, the negative $\delta \phi(k_{h},\nu)$ (c.f. left panel of 
Figure \ref{M1PhaseCoherence}) corresponding to IGWs extend over, along a curved band, to the evanescent region
and beyond into higher frequency domain, over the $k_{h}$ = 4.1 -- 6.8 Mm$^{-1}$ and over $\nu$ beyond 4 mHz.
This seems to indicate that there are IGW-like waves extending beyond the classical gravity wave boundary
expected from the simple dispersion relation given by Equation \ref{dispr}. Interestingly, exactly such a behaviour
is seen in the numerical simulations of IGWs performed by \citet{2020A&A...633A.140V} and as seen in their Figure 5(b),
where they also synthesised spectral lines and derived $\delta \phi(k_{h},\nu)$ from line core intensities of Fe 5576 {\AA} and Fe 5434 {\AA} lines. However, the coherence $C(k_{h},\nu)$ (c.f. right panel of Figure \ref{M1PhaseCoherence}) is 
less than 0.1 in $k_{h}$ = 4.1 -- 5.5 Mm$^{-1}$ $\&$ $\nu >$ 3.5 mHz, 
corresponding to the negative phase observed in the evanescent wave regimes of $\delta \phi$ 
diagrams. Hence the reality of these seemingly high frequency IGWs, in the evanescent region and beyond, 
is doubtful despite their resemblance to the simulation results of \citet{2020A&A...633A.140V}. Nevertheless, there are several factors which can affect the coherence; for example, results of \cite{2020A&A...633A.140V} from 
synthetic observations show that coherence decreases as height separation increases, and we also see such
behaviour for the $I_{c} - I_{uv2}$ intensity pair, which corresponds to higher height (h = 20 -- 430 km). Additionally, they suggested that the angle of wave propagation with respect to  normal for a given wave of a given frequency can also influence the coherence, which is governed by the local
Brunt-V{\"a}is{\"a}l{\"a} frequency ($N$). A wave launched at a particular frequency, without any non-linear interaction, will eventually follow a curved trajectory if the local $N$ changes with height. Therefore, the coherency of the waves propagating between two heights for a given Fourier frequency may locally change over
the field of view \citep{2020A&A...633A.140V}.

\begin{figure*}[ht!]
\centering
\includegraphics[scale=0.42]{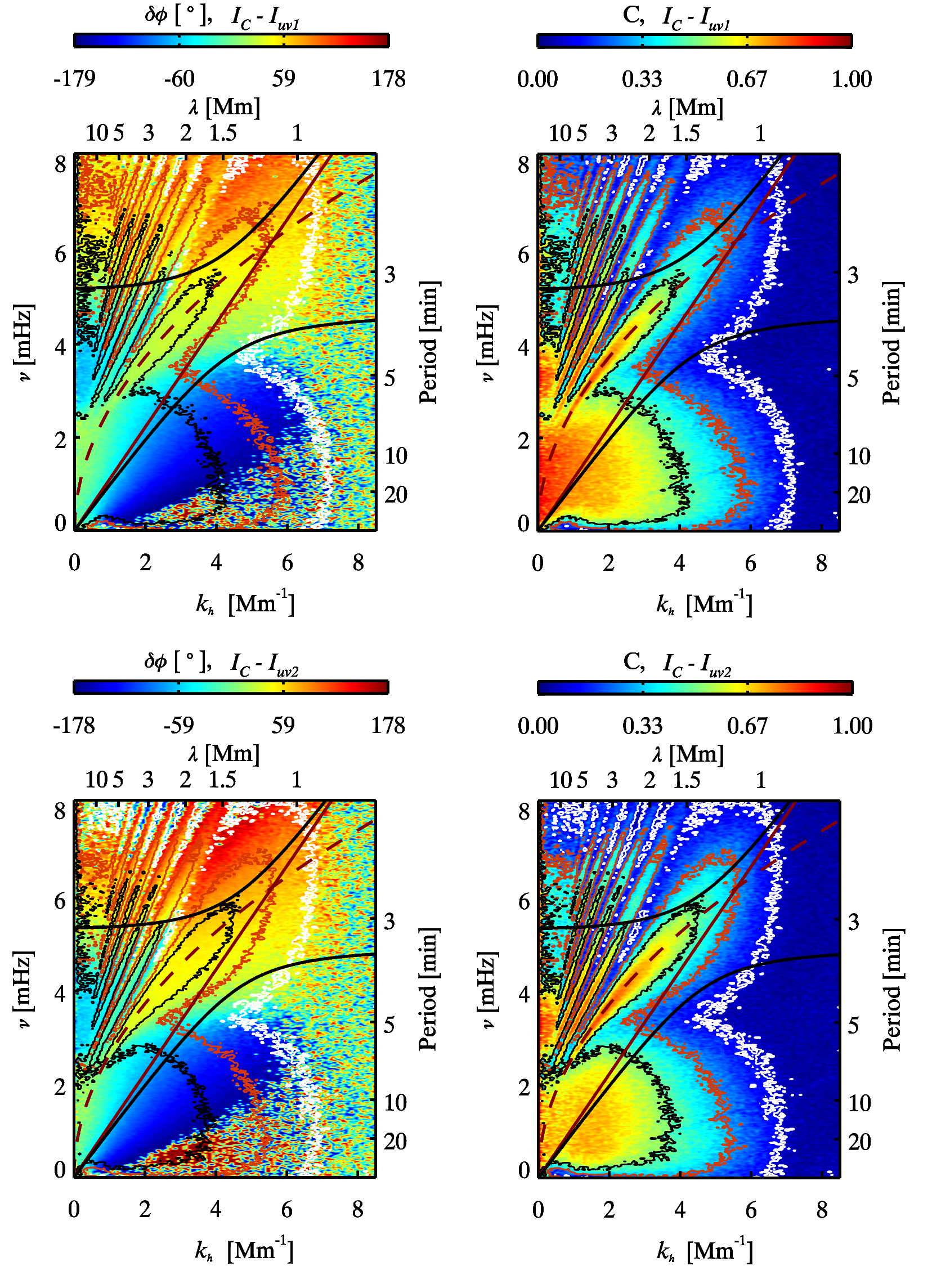}
\caption{\textit{Top panel}: Cross-spectral phase difference, $\delta\phi(k_{h},\nu)$
(\textit{top left panel}), and coherence, C$(k_{h},\nu)$ (\textit{top right panel}), diagrams of
$Q$ region constructed from $I_{c} - I_{uv1}$ pair of photospheric continuum intensity (HMI) and UV
1700 {\AA} channel of AIA, which correspond to 20 -- 360 km above z = 0 in the solar atmosphere.
\textit{Bottom panel}: same as \textit{top panel}, but from $I_{c} - I_{uv2}$ pair of photospheric
continuum intensity (HMI) and UV 1600 {\AA} channel of AIA, which correspond to 20 -- 430 km above z
= 0 in the solar atmosphere. The black solid lines separate vertically propagating waves
($k_{z}^2>0$) from the evanescent ones ($k_{z}^2<0$) at upper height. The dashed red line is the
$f$-mode dispersion curve and solid red line is the Lamb mode. The over plotted black, red and white contours represent the coherence at 0.5, 0.3 and 0.1 levels, respectively.} 
\label{QPhaseCoherence}
\end{figure*}

We show the $\delta \phi$, and $C$ diagrams constructed from another quiet-Sun sub-region labelled as $Q$ within 
a larger region including a small sunspot and neighbouring plage region (c.f. data set 2 shown in Figure 
\ref{fig:hmiblos_dataset2}) in Figure \ref{QPhaseCoherence}. Here, we see the negative $\delta \phi(k_{h},\nu)$
in the gravity wave region extending upto $k_{h}$ = 6.2 Mm$^{-1}$ and also exhibiting 180 deg. wrapping of phase 
at low frequencies (less than 0.8 mHz), especially for the higher height pair $I_{c} - I_{uv2}$ (c.f. bottom left panel of Figure 
\ref{QPhaseCoherence}). Comparing the two quiet-Sun regions $Q$ and $M1$ in left panels of Figures \ref{QPhaseCoherence} 
and \ref{M1PhaseCoherence}, respectively, also reveals significant differences, especially in regard to the curved 
band of negative $\delta \phi(k_{h},\nu)$ that extends through the evanescent region to higher frequencies in the
$M1$ region -- it is absent in the $Q$ region. Furthermore, the $\delta \phi(k_{h},\nu)$ and $C(k_{h},\nu)$ diagrams of $M1$ region shows reduced extent of negative phase and coherence over $k_{h}$ and $\nu$ corresponding to $Q$ region. The observed phase difference of gravity waves in $M1$ region is similar to the earlier estimation by \cite{2003A&A...407..735R} utilizing white light and 1700 {\AA} intensity as shown in their Figure 5, where the overplotted contours at C = 0.5, 0.2 and 0.1 demonstrate that contour of C = 0.5 is extended only upto $k_{h}$ = 1.3 arcsec$^{-1}$, which in Mm$^{-1}$ would be approximately 1.8 Mm$^{-1}$. However, in our analysis the contour of C = 0.5 is extended upto around $k_{h}$ = 2.0 Mm$^{-1}$ (c.f. Figure \ref{M1PhaseCoherence}) and $k_{h}$ = 4.0 Mm$^{-1}$ (c.f. Figure \ref{QPhaseCoherence}). The extent of contour of C = 0.5 in our investigation is better than that earlier reported by  \cite{2003A&A...407..735R}. In the $M1$ region  negative $\delta \phi$ and contour of coherence at $C = 0.1$ is extended only upto $k_{h}$ = 4.1 Mm$^{-1}$ whereas $Q$ region occupies bigger extent than $M1$, upto $k_{h}$ = 6.2 Mm$^{-1}$. This is possibly associated with the slanted propagation of gravity waves in the solar atmosphere. Thus, as we are going away from the disk center, we are probably detecting more and more gravity wave signals compared to that of disk center location. In addition to that, the phase and coherence diagrams constructed from $V - V$ pair i.e. $V_{80} - V_{10}$ over a quiet region ($A$) as depicted in the Figure \ref{IBIS_Sample_Image} are shown in the Figure \ref{Quiet_IBIS}. The gravity wave regime shows the well known negative phase extending upto $k_{h}$ = 6 Mm$^{-1}$ with a very high coherence (c.f. Figure \ref{Quiet_IBIS}). Interestingly, the observed phase difference in the gravity wave regime over a quiet region ($A$) (c.f. Figure \ref{Quiet_IBIS}) is consistent with the earlier phase difference diagram constructed between two height $V - V$ pair estimated within Fe I 7090 {\AA} line as reported in the Figure 1 of \cite{2008ApJ...681L.125S}.

\begin{figure*}[ht!]
\centering
\includegraphics[scale=0.43]{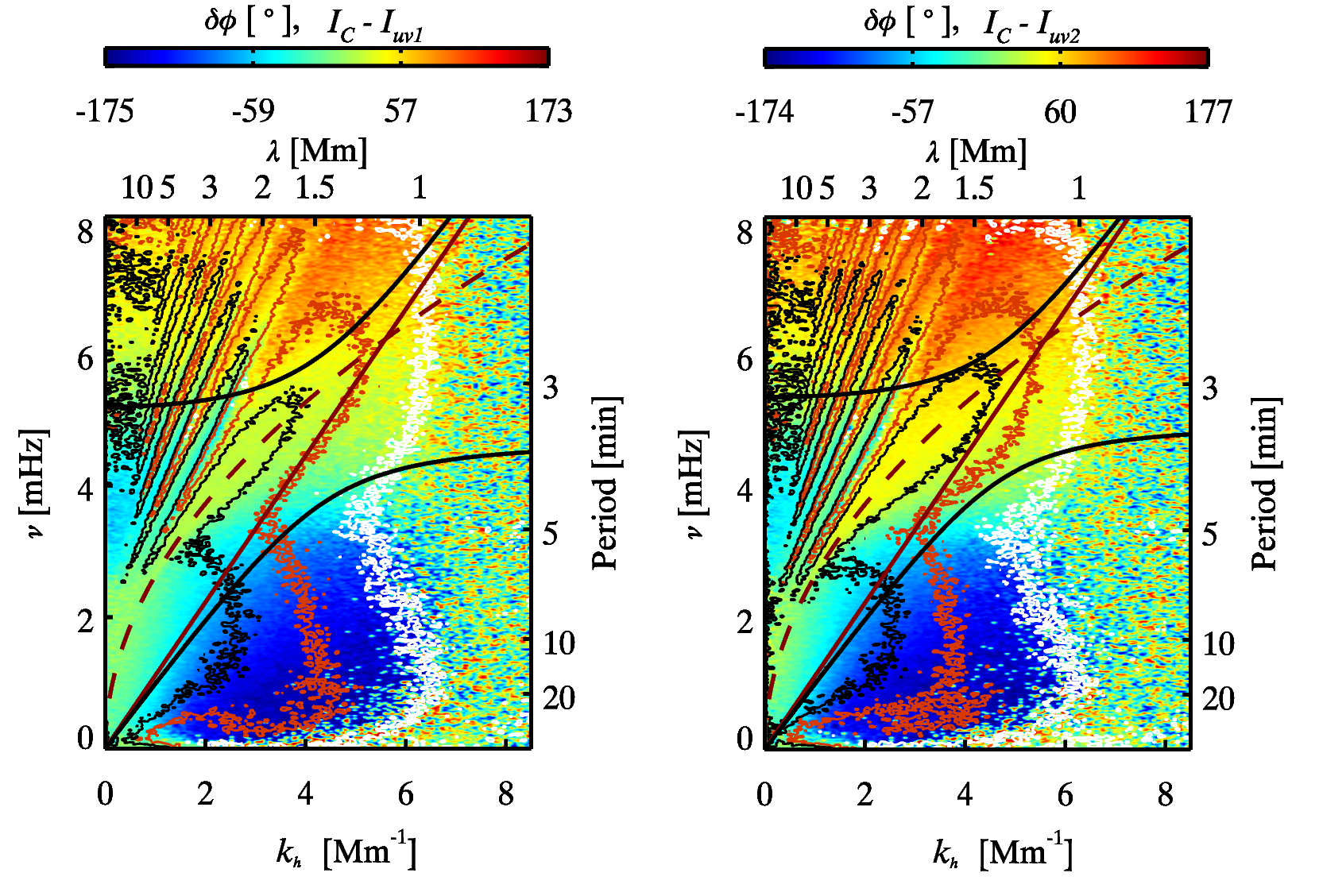}
\caption{Cross-spectral phase difference, $\delta\phi(k_{h},\nu)$ diagrams of $R1$ region as indicated in the Figure \ref{fig:hmiblos_dataset2} constructed from  $I_{c} - I_{uv1}$, and  $I_{c} - I_{uv2}$ intensity pairs, respectively. The over plotted black, red and white contours represent the coherence at 0.5, 0.3 and 0.1 levels, respectively.}
\label{R1PhaseCoherence}
\end{figure*}

\begin{figure*}[ht!]
\centering
\includegraphics[scale=0.43]{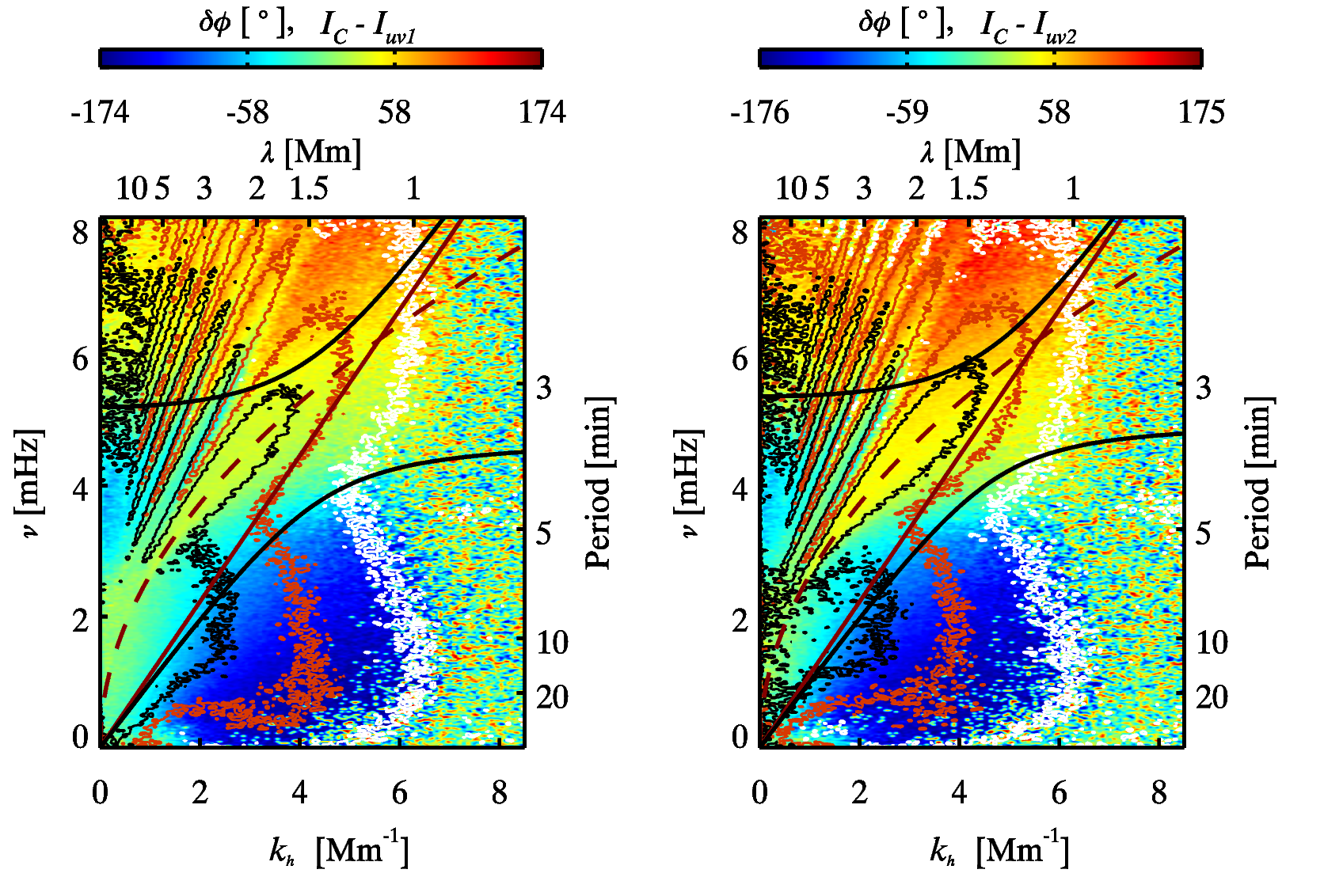}
\caption{Same as Figure \ref{R1PhaseCoherence}, but for $R2$ region as indicated in the Figure
\ref{fig:hmiblos_dataset2}.}
\label{R2_PhaseCoherence}
\end{figure*}

\begin{figure*}[ht!]
\centering
\includegraphics[scale=0.41]{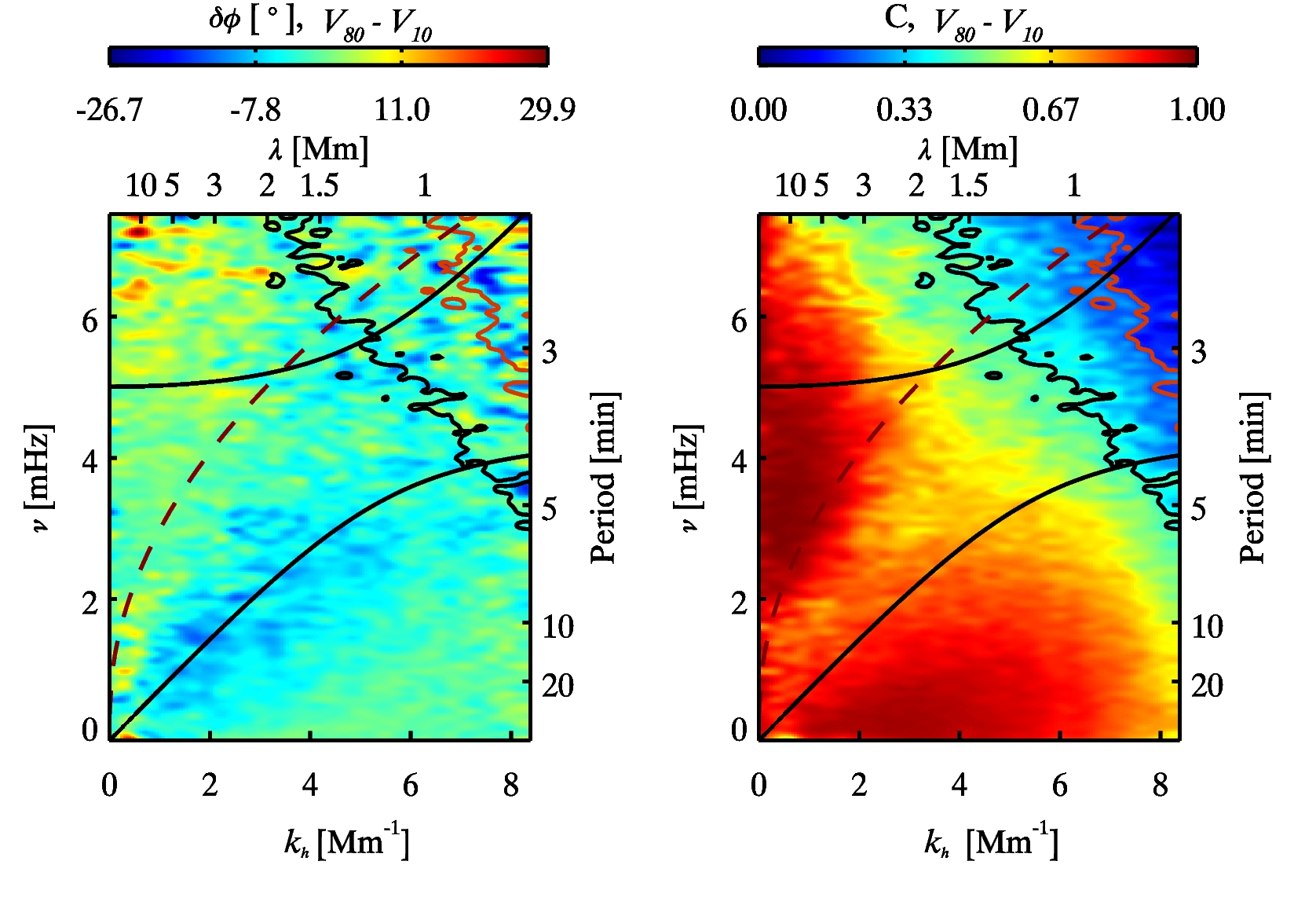}
\caption{Cross-spectral phase difference, $\delta\phi(k_{h},\nu)$
(\textit{left panel}), and coherence, C$(k_{h},\nu)$ (\textit{right panel}), diagrams of quiet ($A$) region constructed from $V_{80} - V_{10}$ velocity pair of photospheric Fe I 6173 {\AA} line observations obtained from the IBIS instrument, which correspond to different heights within 16 -- 302 km above z = 0 in the solar atmosphere. The black solid lines separate vertically propagating waves
($k_{z}^2>0$) from the evanescent ones ($k_{z}^2<0$) at upper height. The dashed red line is the
$f$-mode dispersion curve. The overplotted black, red and white contours represent the coherence at 0.5, 0.3 and 0.1 levels, respectively.}
\label{Quiet_IBIS}
\end{figure*}

\begin{figure*}[ht!]
\centering
\includegraphics[scale=0.41]{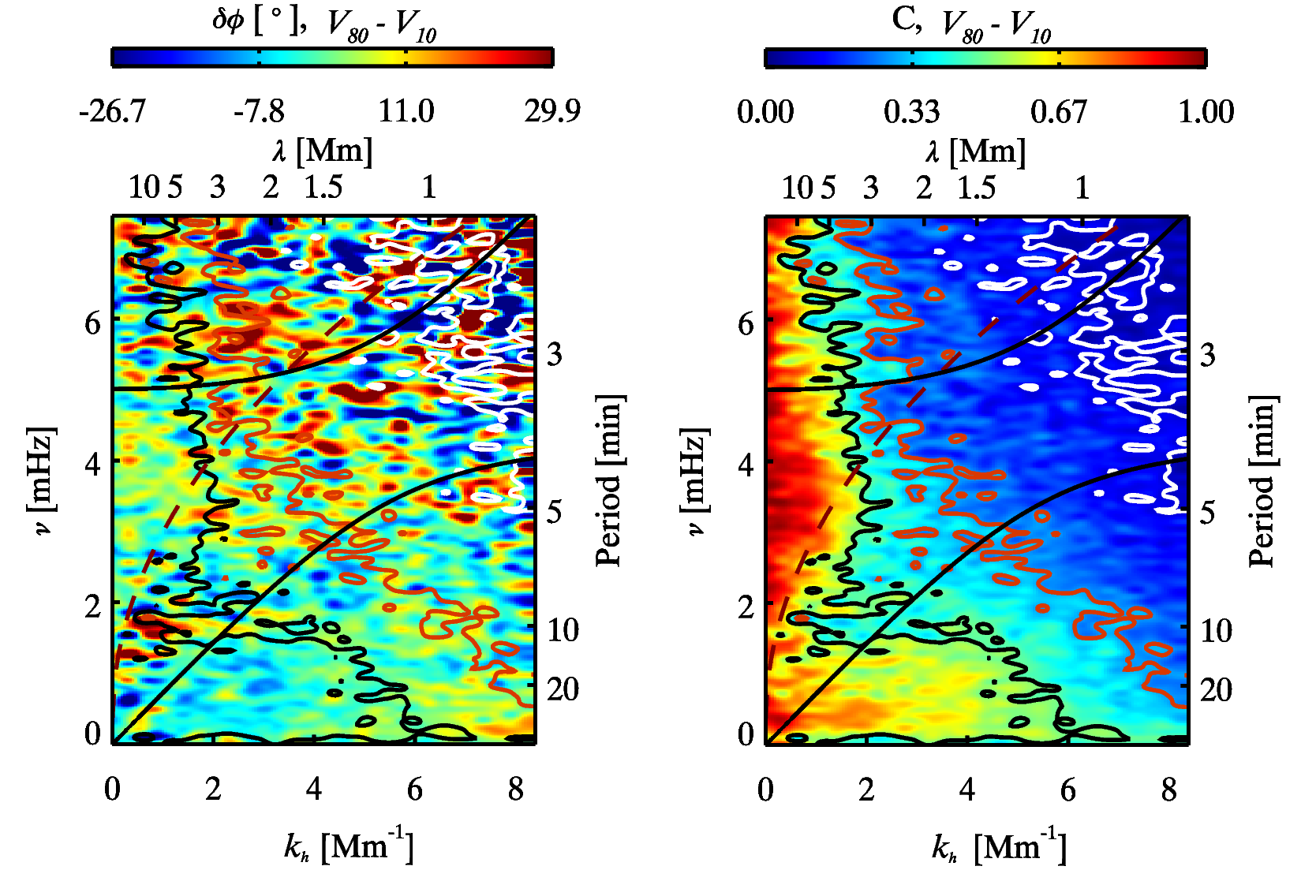}
\caption{Same as Figure \ref{Quiet_IBIS}, but for sunspot ($B$) region as indicated in the Figure \ref{IBIS_Sample_Image}.}
\label{Sunspot_IBIS}
\end{figure*}

\begin{figure*}[ht!]
\centering
\includegraphics[scale=0.23]{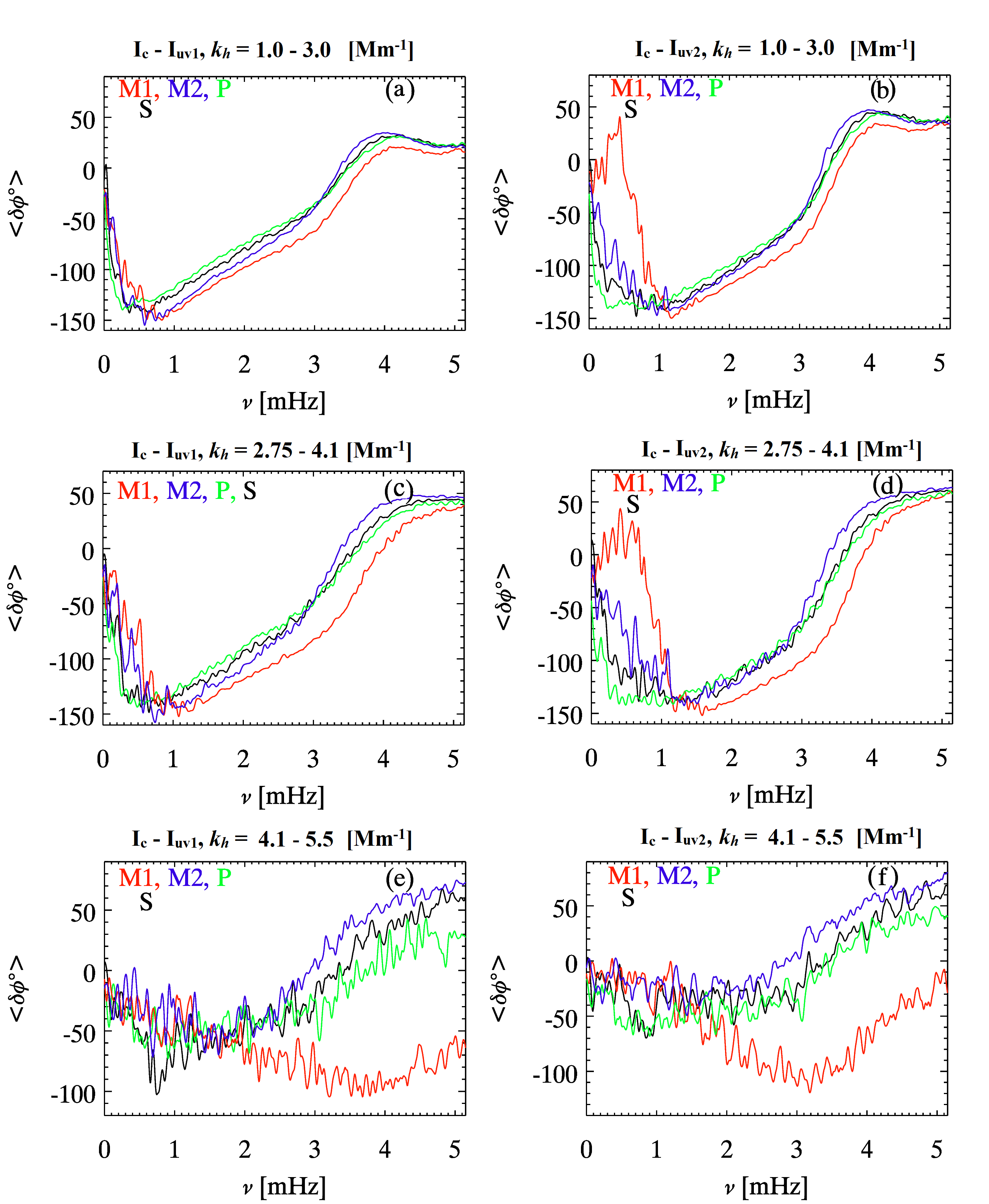}
\caption{Plots indicate average phase over $k_{h}$ = 1.0 -- 3.0, 2.75 -- 4.1, and 4.1 -- 5.5 Mm$^{-1}$, estimated from $I_{c} - I_{uv1}$, and $I_{c} - I_{uv2}$ intensity pairs, over $M1$, $M2$, $P$, and $S$ regions, respectively, as indicated in Figure \ref{fig:hmiblos}.}
\label{Avg_Phase}
\end{figure*}

\begin{figure*}[ht!]
\centering
\includegraphics[scale=0.23]{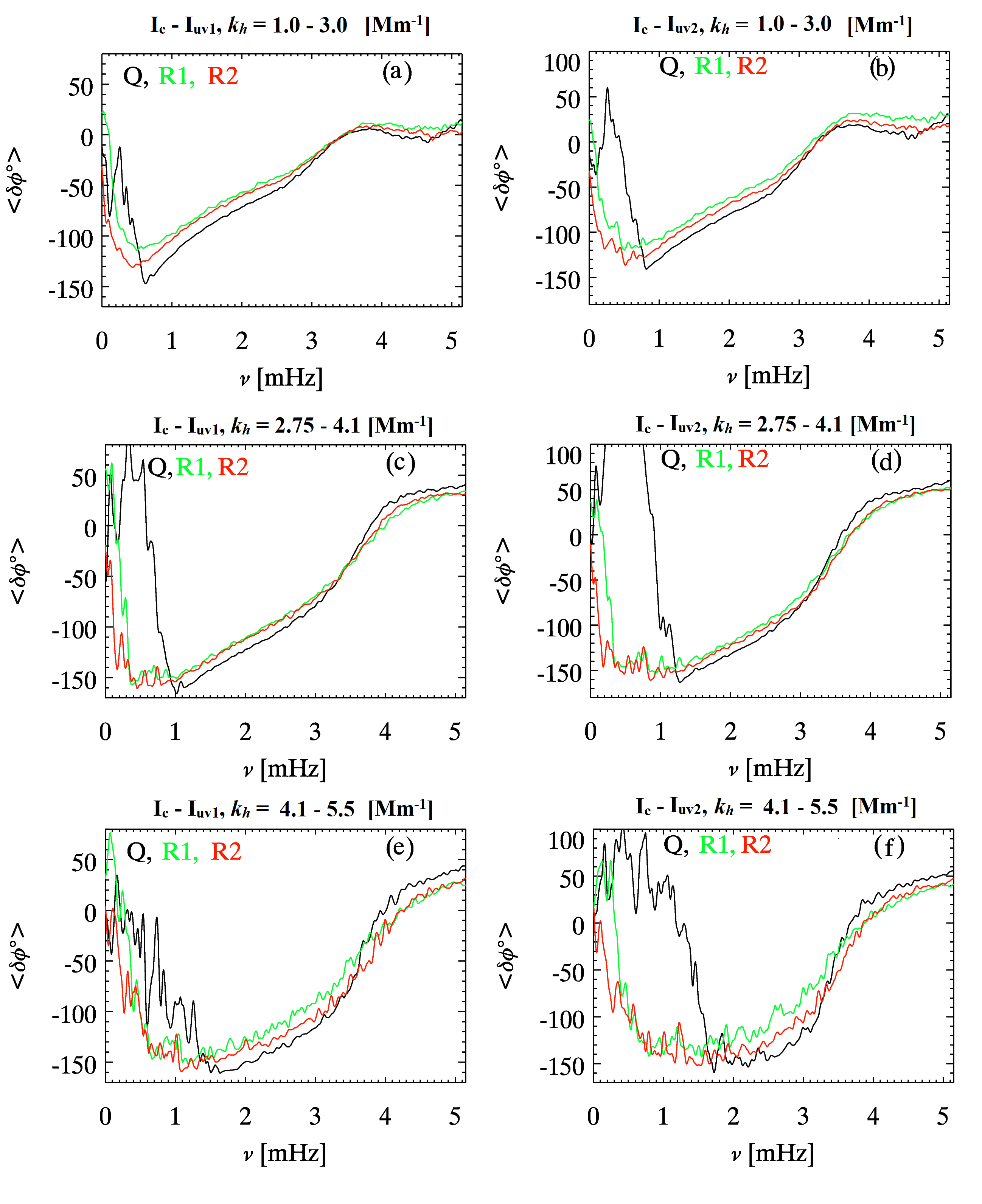}
\caption{Same as Figure \ref{Avg_Phase} but, over $Q$, $R1$, and $R2$ regions, respectively, as indicated in Figure \ref{fig:hmiblos_dataset2}.}
\label{Avg_Phase_dataset2}
\end{figure*}

\begin{figure*}[ht!]
\centering
\includegraphics[scale=0.23]{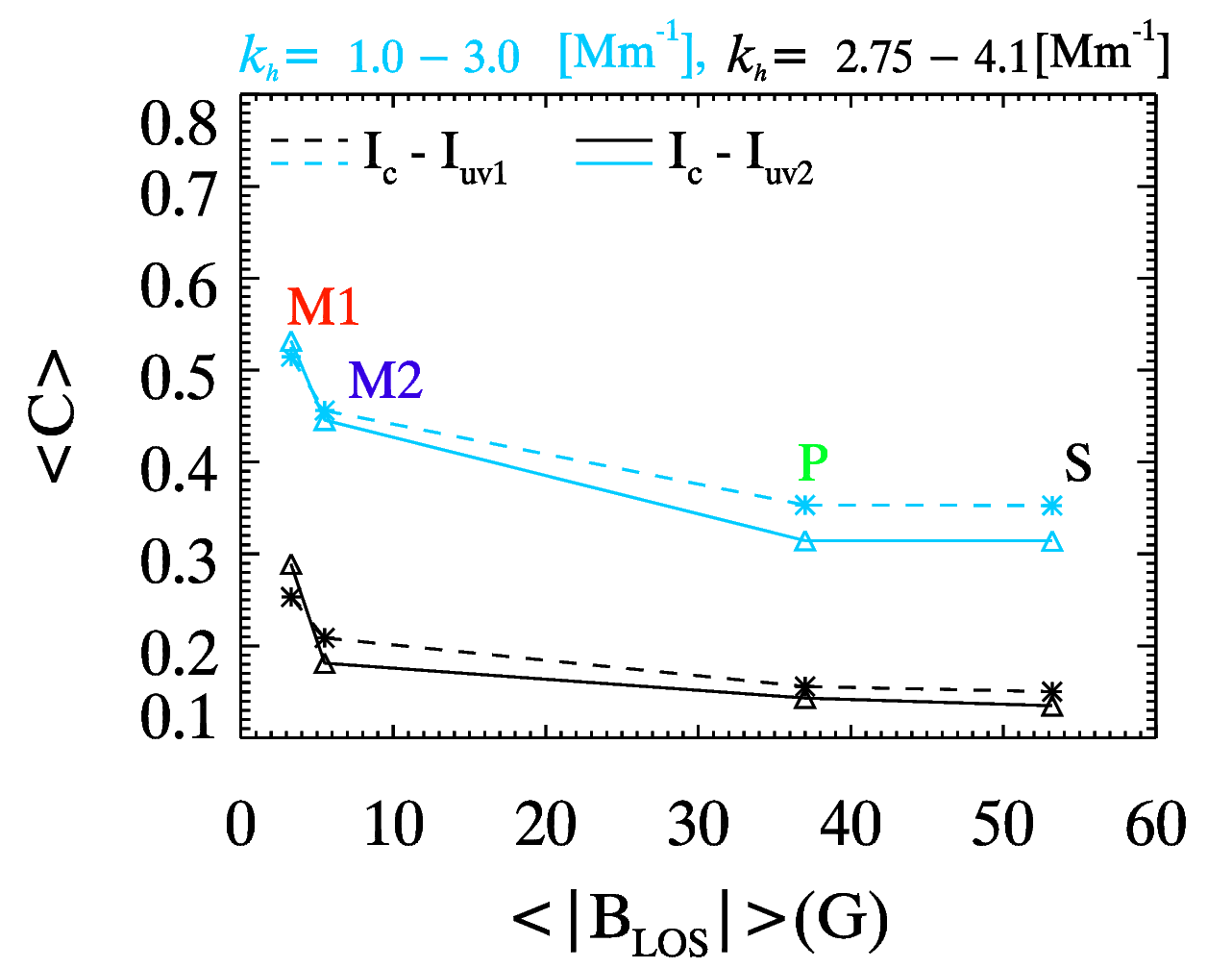}
\caption{Plots show $<C>$ versus $<|B_{LOS}|>$ over $k_{h}$ = 1.0 -- 3.0, 2.75 -- 4.1 Mm$^{-1}$, and $\nu$ = 1 -- 2 mHz, estimated from $I_{c} - I_{uv1}$, and $I_{c} - I_{uv2}$ intensity pairs, over $M1$, $M2$, $P$ and $S$ regions, respectively, as indicated in Figure \ref{fig:hmiblos}.}
\label{Avg_Coh_dataset1}
\end{figure*}

\begin{figure*}[ht!]
\centering
\includegraphics[scale=0.23]{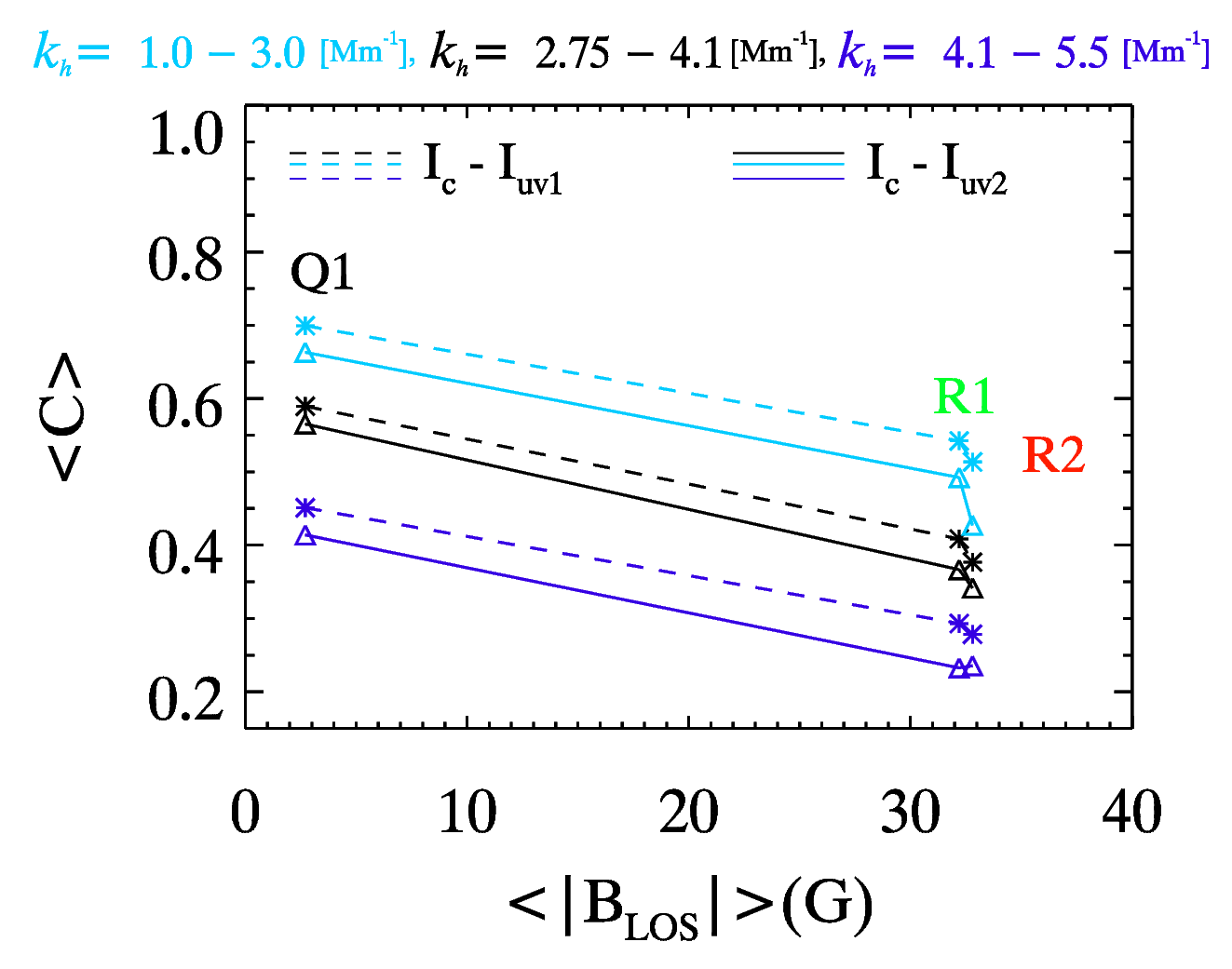}
\caption{Plots show $<C>$ versus $<|B_{LOS}|>$  over $\nu$ = 1 -- 2 mHz, and $k_{h}$ = 1.0 -- 3.0, 2.75 -- 4.1, and 4.1 -- 5.5 Mm$^{-1}$, estimated from $I_{c} - I_{uv1}$, and $I_{c} - I_{uv2}$ intensity pairs, over $Q$, $R1$ and $R2$ regions, respectively, as indicated in Figure \ref{fig:hmiblos_dataset2}.}
\label{Avg_Coh_Dataset2}
\end{figure*}

\begin{figure*}[ht!]
\centering
\includegraphics[width=0.4\textwidth]{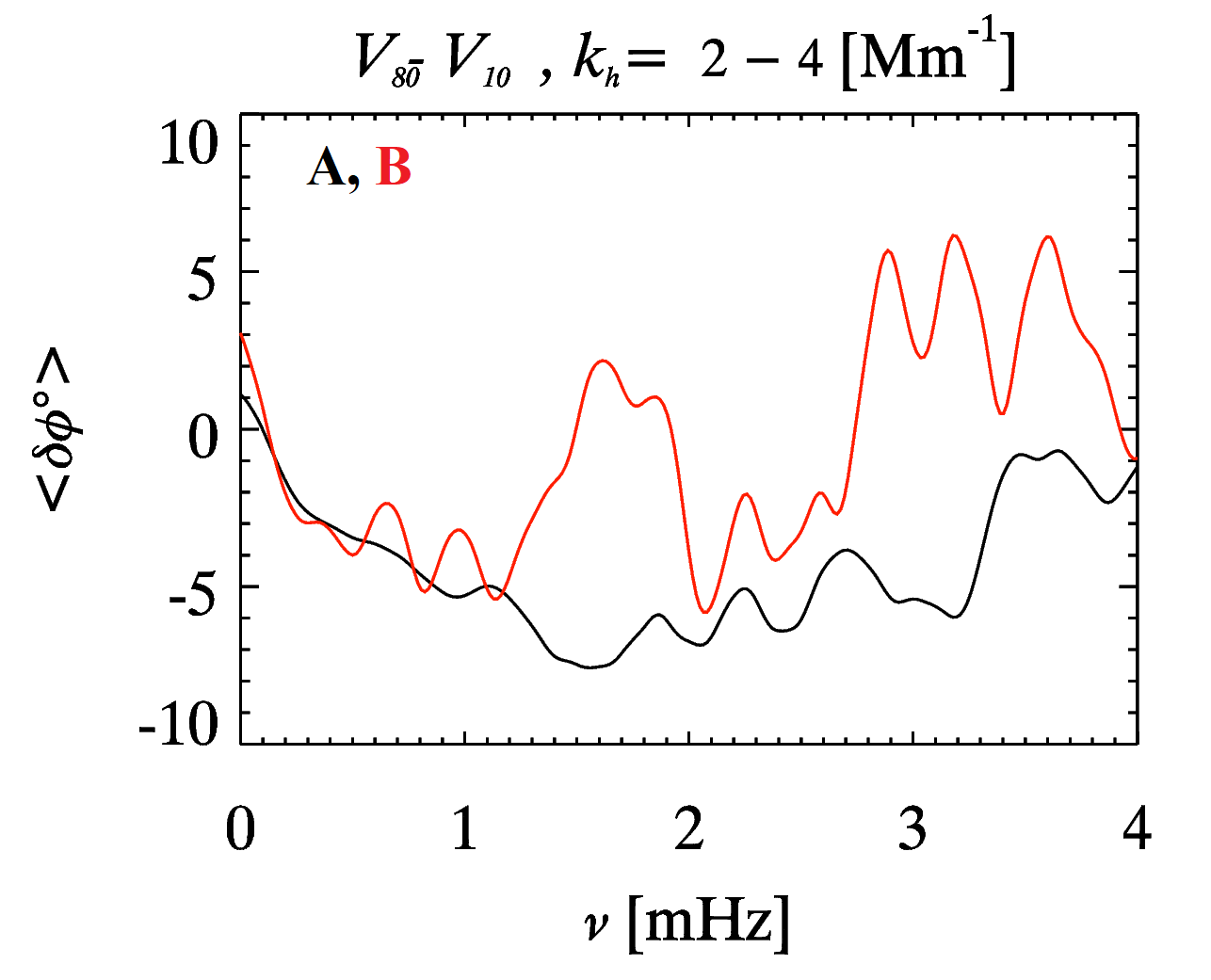}
    \includegraphics[width=0.4\textwidth]{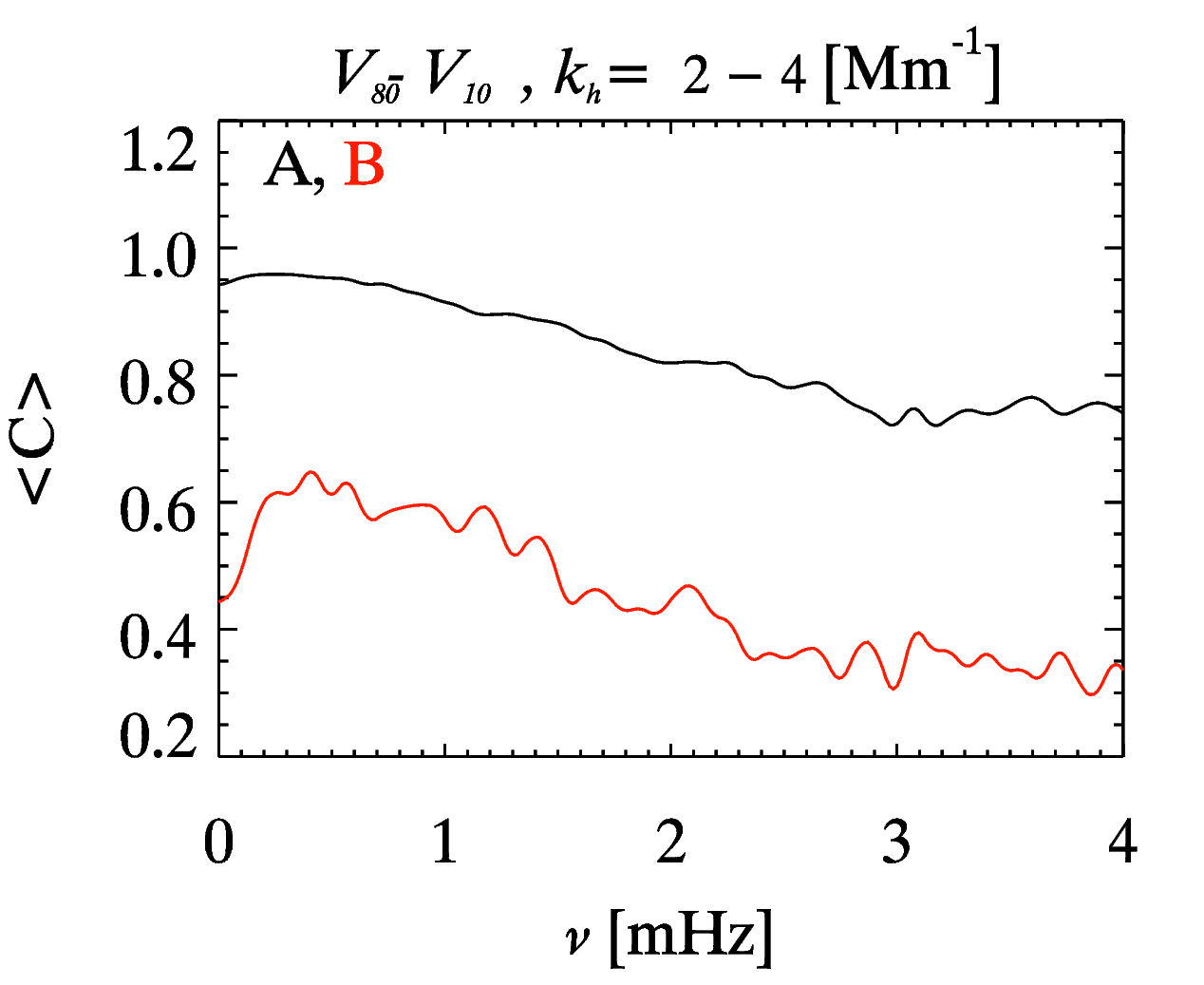}
\caption{Plots indicate average phase (left panel) and coherence (right panel) over $k_{h}$ = 2.0 -- 4.0 Mm$^{-1}$, estimated from $V_{80} - V_{10}$ velocity-velocity pairs, over $A$, and $B$ regions, respectively, as indicated in Figure \ref{IBIS_Sample_Image}.}
\label{Avg_Phase_IBISDataset}
\end{figure*}

\begin{figure*}[ht!]
\centering
\includegraphics[scale=0.28]{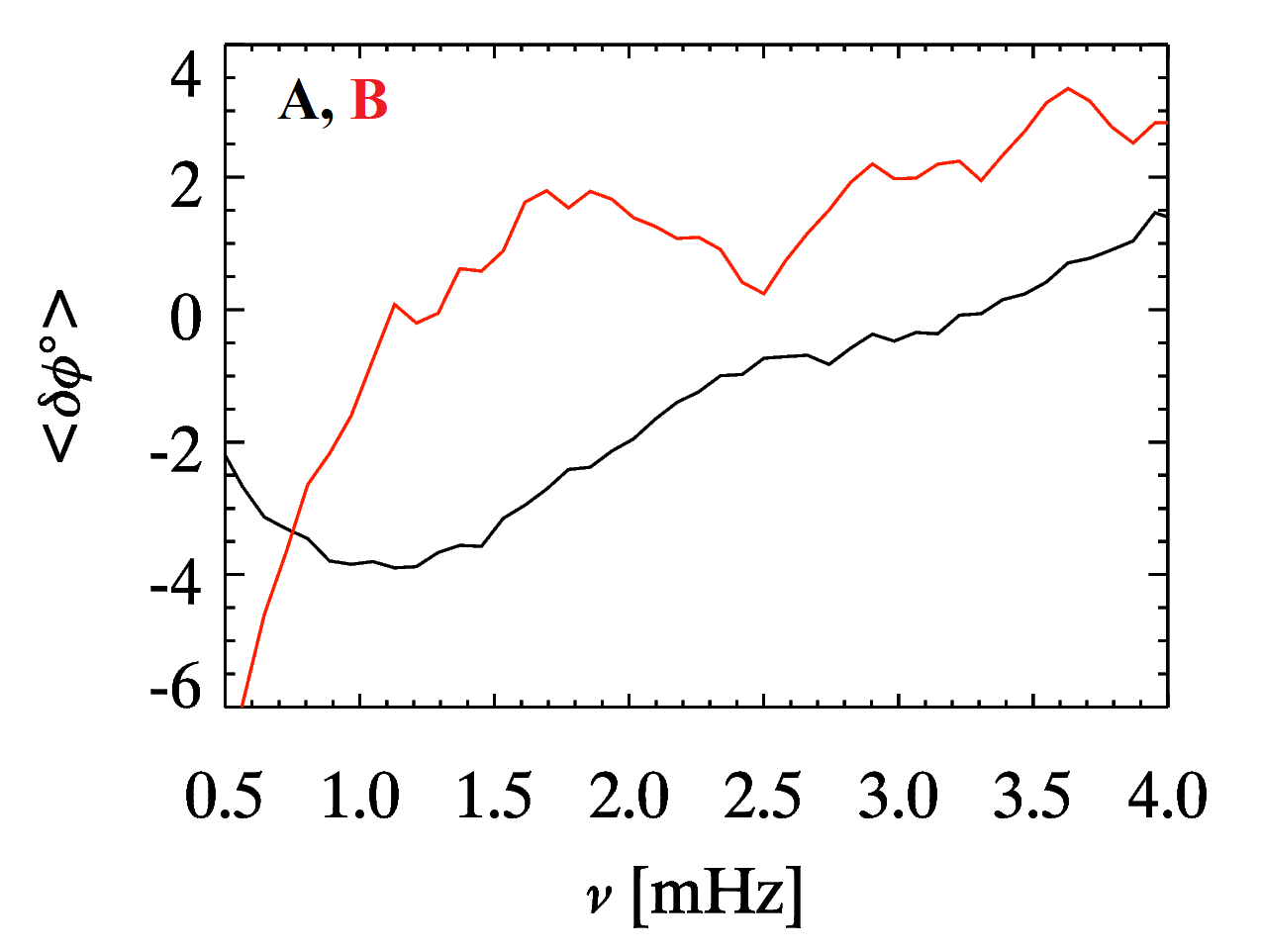}
\caption{Plots indicate average phase estimated over $A$ and $B$ regions from $V_{80} - V_{10}$ pairs from pixel-by-pixel calculation, after removing the acoustic wave regime part above the $f$-mode region in the $k_{h} - \nu$ diagram.}
\label{Avg_Phase_IBISDataset_acoustic_remove}
\end{figure*}

\subsection{Phase Spectra of IGWs in Magnetic Regions and Comparisons with Quiet Regions}

From the data of three large regions (data set 1, 2, and 3), we have several strongly magnetised regions of
different configurations, $P$, $S$, $R1$, $R2$, and $B$ and the $\delta\phi(k_{h},\nu)$ and $C(k_{h},\nu)$ diagrams
of these regions are shown in Figures \ref{Plage_PhaseCoherence}, \ref{Sunspot_PhaseCoherence},
\ref{R1PhaseCoherence}, \ref{R2_PhaseCoherence}, and \ref{Sunspot_IBIS} respectively. The region $M2$ is close to the sunspot
in data set 1 and has one side of the sunspot canopy field covered within it, and it is clear from the
left panels of Figure \ref{M2PhaseCoherence} that the extent of gravity wave regime in the
$M2$ region is reduced over $\nu$ and $k_{h}$ as compared to that seen in the quiet network region
(c.f. Figure \ref{M1PhaseCoherence}): while it occupies $\nu$= 0 -- 4 mHz range extending up to about $k_{h}$ = 5.5 Mm$^{-1}$ in $M1$, the negative phases taper off and become positive over $\nu >$ 3 mHz and
$k_{h} > $ 4.41 Mm$^{-1}$. In their synthetic observations based on numerical simulations, \citet{2020A&A...633A.140V} have 
shown that the magnitude and wavenumber-extent of negative
phase reduces as the magnetic field strength increases. We observe a similar trend in the
$\delta \phi$ diagrams obtained from magnetized regions. In the phase diagrams of $M2$, $P$ and $S$ regions negative
$\delta\phi$ tapers off at approximately $\nu$ = 2.5 mHz and at $k_{h}$ = 4.82 Mm$^{-1}$. Similarly, 
the phase diagrams of $R1$, and $R2$ regions of data set 2 indicate that negative phase is 
extended upto only $\nu$ = 3.0 mHz, and $k_{h}$ = 5.8 Mm$^{-1}$ (c.f. Figures 
\ref{R1PhaseCoherence}, and \ref{R2_PhaseCoherence}); these two regions also exhibit reduced coherence 
as compared to that in the $Q$ region. The above general reducing extents of negative
$\delta \phi$ of IGWs in magnetic regions have to be compared with those observed in the quiet-Sun 
regions $M1$ (c.f. Figure \ref{M1PhaseCoherence}) and $Q$ (c.f. Figure \ref{QPhaseCoherence}), where they extend 
upto about $\nu$ = 4 mHz, and $k_{h}$ = 6.2 Mm$^{-1}$. It is to be noted that the coherence diagrams constructed from $I-I$ pair over quiet regions ($M1$ and $Q$) show that coherence is samll i.e., contour of C at 0.5 extends upto $k_{h}$ = 2 Mm$^{-1}$ and $k_{h}$ = 4 Mm$^{-1}$ as shown in the Figure \ref{M1PhaseCoherence}, and Figure  \ref{QPhaseCoherence}, respectively. Despite the low coherence over the quiet regions (c.f. see extent of contours in Figure \ref{M1PhaseCoherence} and Figure  \ref{QPhaseCoherence}), coherence estimated over magnetized regions ($M2$, $P$, $S$, $R1$ and $R2$) is further reduced (c.f. see extent of contours in Figures \ref{M2PhaseCoherence}, \ref{Plage_PhaseCoherence}, \ref{Sunspot_PhaseCoherence}, \ref{R1PhaseCoherence}, and \ref{R2_PhaseCoherence}). Moreover, the phase and coherence diagrams constructed from $V_{80} - V_{10}$ velocity pair from a high magnetic concentrations over a sunspot ($B$) are shown in the Figure \ref{Sunspot_IBIS}. The $\delta\phi(k_{h},\nu)$ and $C(k_{h},\nu)$ diagrams (c.f. Figure \ref{Sunspot_IBIS}) over $B$  region show that phase and coherence both are reduced in the sunspot over $k_{h}$ and $\nu$ extent, compared to quiet ($A$) region as shown in the Figure \ref{Quiet_IBIS}.\\

In order to better understand the effect of magnetic fields on the IGWs, we have averaged $\delta \phi$ over $k_{h}$ = 1.0 -- 3.0, 2.75 -- 4.1, and 4.1 -- 5.5 Mm$^{-1}$ in $M1$, $M2$, $P$, and $S$ regions of data set 1 and have plotted them in Figure \ref{Avg_Phase}. The left panels [(a), (c), and (e)] correspond to intensity $I_{c} - I_{uv1}$ pair, while right panels [(b), (d), and (f)] are from higher height intensity $I_{c} - I_{uv2}$ pair. We have smoothed $<\delta \phi>$ by taking a smoothing window of 13 point to reduce the noise. In our analysis, the IGWs mostly lie up to $k_{h}$ = 7.0 Mm$^{-1}$, and $\nu$ up to about 3.5 mHz. The panels (a), (b), (c), and (d) of Figure \ref{Avg_Phase}, in the $k_{h}$ = 1.0 -- 3.0, and 2.75 -- 4.1 Mm$^{-1}$ demonstrate that there is a change in the phase of the order of 30 -- 40 and 20 -- 30 degrees in the $\nu$ = 2.5 -- 3 mHz band estimated over $M2$, $P$, and $S$ regions, respectively corresponding to $M1$ region.  Additionally, panels (e) and (f) of Figure \ref{Avg_Phase} indicate that there is a change in the phase of the order of 30 -- 70, and 80 -- 100 degrees in 2.5 -- 3.0 mHz as estimated for the $k_{h}$ = 4.1 -- 5.5 Mm$^{-1}$ wavenumber over $M2$, $P$, and $S$ regions with respect to $M1$ region. Interestingly, we find that the phase is reduced in large amount in the higher wavenumber ($k_{h}$ = 4.1 - 5.5 Mm$^{-1}$) in magnetized regions compared to quiet region, almost changing from negative to positive sign as shown in the panel (e) and (f) of Figure \ref{Avg_Phase}. We also notice that the $<\delta \phi>$ estimated from $M2$ region, which is very close to the sunspot is notably affected in $k_{h}$ = 1.0 -- 3.0, 2.75 -- 4.1, and 4.1 -- 5.5 Mm$^{-1}$ (c.f. Figure \ref{Avg_Phase}) compared to $M1$ region. Similarly, the $<\delta \phi>$ vs $\nu$ plots over $k_{h}$ = 1.0 -- 3.0, 2.75 -- 4.1, and 4.1 -- 5.5 Mm$^{-1}$ for $Q$, $R1$, and $R2$ regions of data set 2 as shown in the Figure \ref{Avg_Phase_dataset2} also demonstrate the reduction in phase over $R1$, and $R2$ regions upto $\nu$ = 4.0 mHz band. The panels (a), (b), (c), and (d) of Figure \ref{Avg_Phase_dataset2} for $k_{h}$ = 1.0 -- 3.0, and 2.75 -- 4.1 Mm$^{-1}$ indicate the change in $<\delta \phi>$ over $\nu$ = 2.5 -- 3 mHz band of the order of 10 -- 20 degrees, while for panels (e) and (f) for $k_{h}$ = 4.1 -- 5.5 Mm$^{-1}$ show the change in $<\delta \phi>$ of the order of 25 -- 50 degree in 2.5 -- 3 mHz band (c.f. Figure \ref{Avg_Phase_dataset2}) concerning to $Q$ region. The bottom panel [(e), and (f)] of Figure \ref{Avg_Phase_dataset2} also shows that phase of gravity waves are reduced in larger amount in the higher wavenumber $k_{h}$ = 4.1 -- 5.5 Mm$^{-1}$ as compared to quiet region ($Q$). Further, it is to be noted that $M2$, and $R1$ regions cover a large amount of looping magnetic fields, hence they are largely horizontal, as can be seen in the AIA 171 {\AA} passband (c.f. right panel of Figures \ref{fig:hmiblos} and \ref{fig:hmiblos_dataset2}), respectively. We also find that suppression of phase is more in these regions ($M2$, and $R1$) as compared to others. \\

Additionally, we also estimate and plot $<\delta \phi>$ over quiet ($A$) and sunspot ($B$) region over $k_{h}$ = 2 -- 4 Mm$^{-1}$ from $V_{80} - V_{10}$ velocity pair as shown in the Figure \ref{Avg_Phase_IBISDataset}. It indicates that $<\delta \phi>$ is positive in the sunspot ($B$) region, while it is negative over the quiet ($A$) region around $\nu$ = 1.5 mHz. Further, we also estimate $<\delta \phi>$ over $A$ and $B$ region from pixel by pixel calculation after removing the acoustic wave regime part from $k_{h} - \nu$ diagram i.e. above the $f$-mode in the diagnostic diagram utilizing a 3D FFT filter. The $<\delta \phi>$ over $A$ and $B$ regions are shown in the Figure \ref{Avg_Phase_IBISDataset_acoustic_remove} shows that there is a change in sign of the $<\delta \phi>$ estimated over sunspot ($B$) region compare to quiet ($A$) region. It is to be noted that we have used velocity $V_{80}$ and $V_{10}$ estimated within the Fe I 6173 {\AA} line at two heights. The formation height of Fe I 6173 {\AA} line is different in quiet and magnetized region i.e. sunspot. In the quiet region, Fe I line forms within $h$ = 16 -- 300 km, whereas in umbra it is around $h$ = 20 -- 270 km above the $\tau_{c}$ = 1. Thus, we estimate the percentage change in the formation height of quiet and sunspot region. There is an approximately 12$\%$ change in the formation height of Fe I line with respect to a quiet region. However, the percentage change in the $<\delta \phi>$  estimated from Figure \ref{Avg_Phase_IBISDataset_acoustic_remove} at $\nu$ = 1.5 mHz are found to be of the order of 114.3$\%$ compare to quiet region. Nevertheless, the $<\delta \phi>$ estimated over a quiet ($A$) and a sunspot ($B$) region from Figure \ref{Avg_Phase_IBISDataset} in $\nu$ = 1.5--3.5 mHz and $k_{h}$ = 2--4 Mm$^{-1}$ are of the order of -5.35 degree and 0.255 degree, respectively. Thus, such a change in $<\delta \phi>$ and change in sign is not possibly due to the lowering in the formation height of Fe I 6173 {\AA} line in sunspot, indicating that it is due to the suppression or reflection of gravity waves in the magnetized regions. The formation height of $I_{uv1}$ and $I_{uv2}$ might be decrease in the magnetized regions. However, the observed percentage change in phase over the highly magnetized regions, are more than 50$\%$--100$\%$ at $\nu$ = 2.0 -- 2.5 mHz and change in sign between quiet and magnetic regions near 3 mHz (c.f. Figure \ref{Avg_Phase}), and their similarities with the phase analysis of $V-V$ spectrum, indicate  that it is not due to change in formation heights of observables used but due to the direct effect of magnetic fields on their propagation. This is because, a sign change in phase shifts between quiet and magnetic regions cannot come from the formation height differences, however big they are.

\subsection{Coherence and Effects of Magnetic Fields}

We analyzed the average coherence on two different wavenumber ranges i.e. $k_{h}$ = 1.0 -- 3.0 Mm$^{-1}$ (pale blue line), and $k_{h}$ = 2.75--4.1 Mm$^{-1}$ (black line) integrated over $\nu$ = 1 -- 2 mHz band from dataset 1, over $M1$, $M2$, $P$, and $S$ regions and plotted them in Figure \ref{Avg_Coh_dataset1}. The plots of $<C>$ versus $<|B_{LOS}|>$ obtained from $I_{c} - I_{uv1}$ (dashed line) and $I_{c} - I_{uv2}$  (solid line) pair as shown in the Figure \ref{Avg_Coh_dataset1} demonstrate that as magnetic field strength increases coherence decreases. From the dataset 2, we also estimate $<C>$ on $k_{h}$ = 1.0 -- 3.0 Mm$^{-1}$ (pale blue line), $k_{h}$ = 2.75--4.1 Mm$^{-1}$ (black line), and $k_{h}$ = 4.1 -- 5.5  Mm$^{-1}$ (blue line) integrated over $\nu$ = 1 -- 2 mHz over $Q1$, $R1$, and $R2$ from $I_{c} - I_{uv1}$ (dashed line) and $I_{c} - I_{uv2}$  (solid line) pairs, which are depicted in the Figure  \ref{Avg_Coh_Dataset2}. These also demonstrate that as magnetic field strength increases coherence decreases. It is to be noted that in $k_{h}$ = 2.75--4.1 Mm$^{-1}$ range in Figure \ref{Avg_Coh_dataset1} and $k_{h}$ = 4.1--5.5 Mm$^{-1}$ regime in the Figure \ref{Avg_Coh_Dataset2} coherence is below 0.5. However, despite the low coherence in quiet regions ($M1$ and $Q1$), coherence is further reduced in magnetized regions. Similar reduction in coherence is also observed in the $<C>$ versus $\nu$ plot constructed over quiet ($A$) and sunspot ($B$) regions as shown in the right panel of Figure \ref{Avg_Phase_IBISDataset}. Figures \ref{Avg_Coh_dataset1}, \ref{Avg_Coh_Dataset2}, and right panel of Figure \ref{Avg_Phase_IBISDataset} suggest that magnetic fields also affect coherence apart from other factors as suggested by \cite{2020A&A...633A.140V}.

\section{Discussion and Conclusions}
\label{sec:final}

The IGWs in the solar atmosphere, which are thought to be generated by the turbulent convection penetrating locally into a stably stratified medium are believed to dissipate their energy by radiative damping just above the solar surface \citep{1967IAUS...28..429L}. Earlier studies \citep{1981ApJ...249..349M, 1982ApJ...263..386M} suggest that IGWs can reach upto the middle chromosphere, before breaking of these waves due to nonlinearities resulting in a complete dissipation. Recent simulations done by \cite{2017ApJ...835..148V}, \cite{2019ApJ...872..166V}, and \citep{2020A&A...633A.140V} have shown that these waves are still present in the higher atmosphere, where the radiative damping time scale is high, and that magnetic fields suppress or scatter IGWs in the solar atmosphere. \\

Our investigations of $k_{h}- \nu$ phases and coherences of IGWs within the photosphere and from photospheric to lower chromospheric height ranges over a varied levels of background magnetic fields and their configurations have brought out clear signatures of reduced extent  of negative phase  and coherence and a change of sign in phases in the gravity wave regime due to the magnetic fields as compared to quiet regions -- as demonstrated in Figures \ref{M1PhaseCoherence}, \ref{M2PhaseCoherence}, \ref{Plage_PhaseCoherence}, \ref{Sunspot_PhaseCoherence}, \ref{QPhaseCoherence}, \ref{R1PhaseCoherence}, \ref{R2_PhaseCoherence}. The above finding from intensity observations is also strengthened from the phase and coherence diagrams estimated from velocity - velocity cross-spectral pairs observed over a quiet ($A$) and a sunspot ($B$) region (c.f. Figure \ref{Quiet_IBIS}, and \ref{Sunspot_IBIS}). Further, from the comparison of average phase in $k_{h}$ = 1.0 -- 3.0, 2.75 -- 4.1, and 4.1 -- 5.5 Mm$^{-1}$ over quiet and magnetic regions from data set 1 and 2 as shown in the Figures \ref{Avg_Phase}, and \ref{Avg_Phase_dataset2}, we find that the phase of IGWs are much reduced in the magnetized regions. Moreover, this effect is more prominent in the higher wavenumber regions i.e. for the waves of lower wavelength ($\lambda$ = 2.175 -- 2.75 Mm) suggesting that these gravity waves are scattered by the background magnetic fields or partially reflecting from the upper atmospheric layers leading to positive phase in the gravity wave regime. The average phase estimated from $V - V$ pair in $k_{h}$ = 2 -- 4 Mm$^{-1}$ (c.f. Figure \ref{Avg_Phase_IBISDataset}) also indicate change in sign at $\nu$ = 1.5 mHz in sunspot ($B$) compare to quiet ($A$) region. Moreover, the average phase estimated from $V - V$ pair from  pixel-by-pixel analysis over quiet ($A$) and sunspot ($B$) regions also demonstrate the change in sign of the phase. Furthermore, the observed percentage change in phase in the sunspot ($B$) compared to quiet ($A$) region is much higher than percentage change in the formation height of Fe I line in sunspot compared to quiet region, strongly favouring the suppression or reflections of gravity waves in the high magnetized region i.e. sunspot (c.f. Figur  \ref{Avg_Phase_IBISDataset_acoustic_remove}). It is to be noted that the observed differences in the $B$ region compared to the $A$ region could be due to lack of excitation of gravity waves over sunspot in the photosphere, but there is still a possibility of scattering/suppression of gravity waves that propagate from the surrounding quiet Sun over the sunspot location. Importantly, we also find a reduction in coherence in the presence of background magnetic fields as shown in the Figures \ref{Avg_Coh_dataset1}, and \ref{Avg_Coh_Dataset2} as well as shown in the right panel of Figure \ref{Avg_Phase_IBISDataset}. Previously, \cite{2008ApJ...681L.125S} showed that the rms wave velocity fluctuations due to IGWs were suppressed at locations of magnetic flux. Here, we have provided further observational evidence in the $k_{h}- \nu$ diagrams for the suppression or scattering or partial reflection of gravity waves in the magnetized regions of the solar atmosphere. \\

In general, the above nature of IGWs in magnetized regions are broadly consistent with the simulation results of \citet{2017ApJ...835..148V} and \cite{2019ApJ...872..166V, 2020A&A...633A.140V} suggesting the suppression of gravity waves in magnetized regions. This led to identifying our reported differences between quiet and magnetic regions as observational evidences for such influences of magnetic fields.
We anticipate that several of our analyses and findings reported here will be examined in more detail in the near future containing bigger FOV, involving coordinated simultaneous multi-height observations of photosphere and chromosphere utilising Dopplergrams data from the newer Daniel K Inouye Solar Telescope (DKIST; \cite{2020SoPh..295..172R}) (National Solar Observatory USA) facility along with MAST \citep{2009ASPC..405..461M, 2017CSci..113..686V} and HMI, AIA onboard SDO spacecraft.\\

\section*{Acknowledgments}
We acknowledge the use of data from the HMI and AIA instruments onboard the {\em Solar Dynamics Observatory} spacecraft of NASA. We are thankful to SDO team for their open data policy. The SDO is NASA's mission under the Living With a Star (LWS) program. At the time IBIS observation was acquired, Dunn Solar Telescope at Sacramento Peak, New Mexico, was operated by National Solar Observatory (NSO). NSO is managed by the Association of Universities for Research in Astronomy (AURA) Inc. under a cooperative agreement with the National Science Foundation. The Research work being carried out at Udaipur Solar Observatory, Physical Research Laboratory is supported by the Department of Space, Govt. of India. S.P.R. acknowledges support from the Science and Engineering Research Board (SERB, Govt. of India) grant CRG/2019/003786. We thank two  anonymous referees for critical and constructive comments and suggestions which significantly improved the presentation and discussion of results in this paper. 

\bibliographystyle{jasr-model5-names}
\biboptions{authoryear}
\bibliography{refs}

\end{document}